\documentclass[a4paper,11pt]{article}

\usepackage{jheppub} 
\usepackage{tikz}

\renewcommand{\d}{\text{d}}
\newcommand{\tr}{\text{Tr}}
\newcommand{\be}{\begin{equation}}
\newcommand{\ee}{\end{equation}}
\newcommand{\bea}{\begin{eqnarray}}
\newcommand{\eea}{\end{eqnarray}}

\title{\boldmath More on $T \overline{T}$-like  deformations in higher dimensions}

\author[a,b]{Nicol\`o Brizio,}
\author[a,b]{Moritz Kade,}
\author[b,c,1]{Alessandro Sfondrini,\note{On leave from the University of Padova, Italy.}}
\author[b,a]{Dmitri P. Sorokin}

\affiliation[a]{Dipartimento di Fisica e Astronomia, Universit\`a degli Studi di Padova, Via Marzolo 8, 35131 Padova, Italy}
\affiliation[b]{INFN, Sezione di Padova, Via Marzolo 8, 35131 Padova, Italy}
\affiliation[c]{School of Mathematics, University of Birmingham, Edgbaston B15 2TT, UK}

\emailAdd{nicolo.brizio@unipd.it}
\emailAdd{moritztill.kade@unipd.it}
\emailAdd{a.sfondrini@bham.ac.uk}
\emailAdd{dmitri.sorokin@pd.infn.it}

\abstract{
We investigate several possible generalisations of $T\overline{T}$ deformations to three- and higher-dimensional field theories. Starting from the two-dimensional $T\overline{T}$ flow, we work out its higher-dimensional uplift, which results in a non-local and non-isotropic three-dimensional theory.
Starting instead from the relation between the Nambu--Goto action and  $T\overline{T}$ in $d=2$, we study the flow equation obeyed by the Dirac--Nambu--Goto actions in $d>2$ dimensions, written in terms of the stress-energy tensor only. Similarly, we derive the stress-tensor flow obeyed by the Born--Infeld actions in $d$ dimensions and by the Dirac--Born--Infeld actions in $d=2$ and $d=3$.
}

\begin{document}
\makeatletter
\gdef\@fpheader{}
\makeatother
\maketitle
\flushbottom

\section{Introduction}\label{intro}
A rich class of deformations of two-dimensional field theories is given by current--current deformations. Starting from a seed action~$\mathcal{S}_{0}$ and two conserved currents $J^\mu$ and $K^\mu$, one defines a classical flow for the one-parameter family of actions $\mathcal{S}_\lambda$
\begin{equation}\label{first_flow}
    \frac{\partial\mathcal{S}_\lambda}{\partial\lambda}=
    \int\d^2x\ \mathcal{O}(x)\,,\qquad
    \mathcal{O}(x)=-\frac12\,\epsilon_{\mu\nu}J^{\mu}(x)\,K^{\nu}(x)\,,\qquad
    \partial_\mu J^{\mu}= \partial_\mu K^{\mu}=0\,.
\end{equation}
Remarkably, this gives a well-defined operator at the quantum level~\cite{Smirnov:2016lqw}, so that we may promote the flow labeled by~$\lambda$ to a family of quantum field theories. A type of deformation which stands out for its universality is the so-called $T\overline T$ one~\cite{Cavaglia:2016oda,Smirnov:2016lqw}, in which one sets $J^\mu=T^{0\mu}$ and $K^\mu=T^{1\mu}$ in terms of the stress-energy tensor~$T^{\mu\nu}$. The resulting operator%
\footnote{In what follows we will use $\text{det}[M_{\mu}{}^\nu]$ to denote the determinant of a matrix, while $\text{det}[M_{\mu\nu}]$ will refer to the determinant of a tensor $M_{\mu\nu}$, obtained from $M_{\mu}{}^{\nu}$ by lowering its second index with a metric of Lorentz signature. We use the convention in which the signature of the Minkowski metric is almost plus.}
\begin{equation}\label{ttintro}
  \mathcal{O}_{T\overline T} = \frac{1}{4}\left[T^{\mu\nu}T_{\mu\nu}-(T_\mu{}^\mu)^2 \right] =  -\frac{1}{2}\det T_{\mu}{}^{\nu}=\frac{1}{2}\det T_{\mu\nu}
\end{equation}
is a Lorentz scalar. It can be defined without any reference to the internal features of the theory.%
\footnote{Another notable example of current--current deformations arises from a theory with a conserved irrotational current, $\partial_\mu J^\mu=\varepsilon_{\mu\nu}\partial^\nu J^\mu=0$, by setting $K^\mu=\epsilon^{\mu\nu}J_\nu$. These are the $J\overline{J}$ deformations which are important in particular in the context of conformal field theory, see e.g.~\cite{Borsato:2023dis}.
More possibilities open up if we relax Lorentz invariance, such as $J\overline{T}$ deformations~\cite{Guica:2017lia}.}
It is immediate to see that these deformations are \textit{irrelevant} in the renormalisation group sense, which may be a reason why they were overlooked until relatively recently. Nonetheless, the $T\overline T$ deformations have attracted a great deal of attention because of their remarkable properties, see e.g.~\cite{Jiang:2019epa,He:2025ppz} and references therein. In particular:
\begin{itemize}
    \item Such deformations preserve integrability, if present in the seed theory~\cite{Cavaglia:2016oda, Smirnov:2016lqw}. Indeed, the $T\overline T$ operator was originally introduced in the context of integrable renormalisation group flows~\cite{Zamolodchikov:2004ce}, see also~\cite{Zamolodchikov:1991vx, Mussardo:1999aj}.
    \item They act in a simple way on the S-matrix of a two-dimensional QFT, by ``dressing'' it with a specific scalar Castillejo-Dalitz-Dyson (CDD) factor. Intriguingly, this CDD factor is the same as appears in the description of effective string theories~\cite{Dubovsky:2012sh, Dubovsky:2012wk} and on the worldsheet of integrable string models in the light-cone gauge~\cite{Arutyunov:2005hd,Baggio:2018gct}.
    \item It is possible to express any energy level of the deformed theory in terms of the corresponding level of the undeformed model in a closed form. The deformed spectrum takes a Hagedorn form~\cite{Hagedorn:1965st}, which yields another connection with string theory.
    \item Perhaps the most striking relation to string theory (and the first one to be noticed~\cite{Cavaglia:2016oda}) is that the $T\overline{T}$-deformed action of $N$ free scalars is closely related to the Nambu--Goto action in $(N+2)$-dimensional target space-time in a static gauge. More precisely, the $T\overline{T}$ deformation can be related to the uniform lightcone gauge~\cite{Baggio:2018gct}, something which can be also used to efficiently and explicitly construct deformed actions~\cite{Frolov:2019nrr,Frolov:2019xzi}.
    \item The deformation can be interpreted as coupling the original model to two-dimensional gravity~\cite{Dubovsky_2017, Cardy:2018sdv, Tolley:2019nmm, Caputa:2020lpa} or to a field-dependent background metric \cite{Conti:2018tca}.
    \item Finally, $T\overline{T}$ deformations may also be interpreted holographically in the context of the $\text{AdS}_3/\text{CFT}_2$ correspondence, where one thinks of deforming the dual $\text{CFT}_2$, see~\cite{McGough:2016lol, Giveon:2017nie, Kraus:2018xrn, Guica:2019nzm}.
\end{itemize}

When one tries to extend the notion of the $T\bar{T}$ deformation to space-time dimensions higher than two, one cannot expect all the features listed above to carry over. For instance, the relation with integrability is quite natural in two dimensions, but will probably be the hardest to generalise to higher dimensions --- integrable QFTs are quite rare and constrained in higher dimensions to begin with. More generally, finding a universal definition of the deforming operator that is consistent at the \textit{quantum} level and applies to any (or most) interacting QFTs seems like a very difficult task in higher dimensions.%
\footnote{In $d=2$ it is possible to prove~\cite{Smirnov:2016lqw} that $\mathcal{O}(x,y)=\epsilon_{\mu\nu}J^\mu(x)K^\nu(y)$ is regular as $y\to x$, but in $d>2$ we would expect short-distance singularities to appear, at least for generic interacting theories.
}
A handle on the quantum definition of the deformation (or perhaps a shortcut) is provided by the AdS/CFT correspondence, which allows one to infer the properties of a higher-dimensional generalisation of $T\overline{T}$ from an analysis of the (classical) gravitational dual. This strategy was adopted in~\cite{Taylor:2018xcy}, where it was proposed that~\eqref{ttintro} may be generalised as
\begin{equation}
    \label{eq:TTbarTaylor}
    \mathcal{O}_{T\overline T,d} = \frac{1}{4}\left[T^{\mu\nu}T_{\mu\nu}-\frac{1}{d-1}(T_\mu{}^\mu)^2\right]\,.
\end{equation}
Operators of this type were further studied in~\cite{Conti:2018jho,Khoeini-Moghaddam:2020ymm,Conti:2022egv} where the coefficient $1/(d-1)$ was further generalised to some function~$r(d)$.
Different considerations may lead to different generalisations. For instance, a stochastic interpretation of $T\overline{T}$  deformations lead to an operator of the form $\mathcal{O}_{T\overline T,d}=\tfrac{1}{2}|\text{det} T_{\mu\nu}|^{\frac 1{d-1}}$. This, of course, has drastically different properties (and scaling dimension) than~\eqref{eq:TTbarTaylor}. Other generalisations of $T\overline{T}$ deformations to $d\geq2$ were considered e.g. in~\cite{Bonelli:2018kik,Conti:2018jho,Babaei-Aghbolagh:2020kjg,Khoeini-Moghaddam:2020ymm,Rodriguez:2021tcz,Tempo:2022ndz,Babaei-Aghbolagh:2022uij,Ferko:2022iru,Conti:2022egv,Ferko:2022cix,Borsato:2022tmu,Ferko:2023sps, Ferko:2023wyi,Morone:2024ffm,Ferko:2024zth,Babaei-Aghbolagh:2024hti,Babaei-Aghbolagh:2024uqp,Babaei-Aghbolagh:2025uoz,Li:2025lpa,Nix:2025plr}.

The aim of this paper is to revisit the problem of extending $T\overline{T}$ deformations to higher dimensions from two independent and rather different points of view.

\begin{figure}[t]
\centering
    \begin{tikzpicture}
\node (free3) at (-5,2) {$\displaystyle \mathcal{S}_0=\int\d^2x\int\limits_0^{2\pi R}\d y\;\mathcal{L}_{0}(\Phi)$};
\node (free2) at (-5,-2) {$\displaystyle \mathcal{S}_0=\int\d^2x\;\mathcal{L}_{0}(\Phi_n;R)$};
\node (TT2) at (5,-2) {$\displaystyle \mathcal{S}_\lambda=\int\d^2x\;\mathcal{L}_{\lambda}(\Phi_n;R)$};
\node (TT3) at (5,2) {$\displaystyle \mathcal{S}_\lambda=\int\d^2x\int\limits_0^{2\pi R}\d y\;\mathcal{L}_{\lambda}(\Phi;R)$};
\draw[->,thick] (free2) -- (TT2) node [pos=0.5,above] {$\partial_\lambda\mathcal{L}_\lambda(\Phi_n;R)=\tfrac{1}{2}\,\text{det} [T_{\mu\nu}] $};
\draw[->,thick] (free3) -- (free2) node [pos=0.5,sloped,above] {compactify};
\draw[->,thick] (TT3) -- (TT2) node [pos=0.5,sloped,above] {compactify};
\draw[->,thick] (free3) -- (TT3) node [pos=0.5,above] {$\partial_\lambda\mathcal{L}_\lambda(\Phi)=\mathcal{O}_{\text{non-loc}}$};
    \end{tikzpicture}
    \caption{Starting from a $d=3$ model in the top left corner (in the simplest case, a free theory of a single massless boson~$\Phi$), we can obtain a two-dimensional theory with infinitely many fields $\Phi_n$, $n\in\mathbb{Z}$ (bottom left). This can be formally deformed by the usual $T\overline{T}$ flow in $d=2$ (bottom right). We can then reverse-engineer the $d=3$ interacting theory (top right) and the flow that would give the bottom right model by compactification.}
    \label{fig:TT-diagram}
\end{figure}

Our first approach builds on the discussion of~\cite{Seibold:2023zkz}. There, the authors were interested in the dynamics of three-dimensional membranes and in their integrability (see also~\cite{Seibold:2024oyr}). Among other things, they dimensionally reduced the three-dimensional membrane action on a circle, obtaining a theory for the resulting infinitely many Kaluza--Klein (KK) modes, which turned out not to be integrable. A natural question is what would happen if one takes a free (and hence, integrable) theory in $d=3$, compactifies down to $d=2$, performs the usual $T\overline{T}$ deformation~\eqref{ttintro}, and tries to uplift back to $d=3$. It was suggested in~\cite{Seibold:2023zkz} that this would yield a non-local model. This is summarised in Figure~\ref{fig:TT-diagram}.
Here we further explore this idea and work out the form of the resulting $d=3$ deformed theory in terms of a formal power series in the deformation parameter, which, order by order, is less and less local.
While this deformation has the closest relation to integrability, it is quite unwieldy, which motivates us to consider alternative approaches.

One of such approaches can be to study the flow of the classical Lagrangians as differential equations for higher-dimensional models, which generalise some of the above properties of $2d$ $T\overline{T}$~\cite{Bonelli:2018kik}.
Perhaps the most natural candidate is the Dirac--Nambu--Goto (DNG) action for a $p$-brane with a $d=p+1$ dimensional worldvolume propagating in a $D$-dimensional target space, which generalises the case of the string~\cite{Dirac:1962iy, Goto:1971ce, Nambu:1974zg}. Another is the case of the Born--Infeld (BI) type actions, which describe $d$-dimensional non-linear electrodynamics, whose relation with $T\overline{T}$ was noticed for $d=4$ already in~\cite{Conti:2018jho}.
Starting from such $\lambda$-dependent actions (in the case of the $p$-brane, $\lambda$ is the inverse tension), we find that, for any number of space-time dimensions, the flow equation can be written exclusively in terms of the stress-energy tensor (and may contain an explicit dependence on $\lambda$). This was already observed in~\cite{Blair:2024aqz} for the $p$-brane action.%
\footnote{We thank Gabriele Tartaglino-Mazzucchelli for bringing this work to our attention.}
Our treatment also applies to the case of $d$-dimensional BI and, in $d=2$ and $d=3$, to the Dirac--Born--Infeld (DBI) action, as we will see. It is worth noting that in a given dimension, along each of these flows, there may hold particular identities for the stress-energy tensor so that in general the deforming operator cannot be written in a unique way.

The rest of the paper is organised as follows. In section~\ref{sec:compact_p2}, we discuss the dimensional uplift of two-dimensional $T\overline T$ deformations to $d=3$, following~\cite{Seibold:2023zkz}. section~\ref{DNG} reviews the universal form of the stress-tensor operator appearing in the flow equation for $p$-branes~\cite{Blair:2024aqz}. We then solve the flow equation for seed theories with a scalar-field potential and obtain the corresponding novel deformed actions. In section~\ref{BI} we derive a universal expression for the stress-tensor operator of $d$-dimensional BI theory. We then turn to the DBI case in section~\ref{split}. In particular, we focus on the two-dimensional DBI theory and show that, for a single scalar field, it obeys a genuine stress-tensor flow (i.e., one which only depends on $T_{\mu\nu}$, without the explicit dependence on the BI field strength found in~\cite{Blair:2024aqz}). We also review the derivation of a stress-tensor flow equation of the three-dimensional DBI theory carried out in \cite{Blair:2024aqz}. Finally, we show that the dimensional reduction of the BI flow equation from $d=4$ to $d=3$ yields the DBI flow equation --- just as it happens for the actions. In section~\ref{conclusions} we summarise our results and outline possible directions for future work.

\section{Uplifting the  \texorpdfstring{$T\overline{T}$}{TTbar} flow to higher dimensions}
\label{sec:compact_p2}
As we have mentioned in the introduction, a natural way to define a higher-dimensional $T\overline{T}$ flow in $d>2$ is to compactify a seed model of interest down to $d=2$, work out the ``usual''  $T\overline{T}$-like flow therein, and uplift back to higher dimensions. One advantage of this approach is that it should preserve the integrability of the original model. While there are some intriguing classically integrable models in $d>2$, here we would like to test this idea in the simplest possible setup, i.e., for a free scalar theory in $d=3$.
As we will see, this case is already quite non-trivial as it involves an infinite tower of KK modes (see e.g. \cite{Bailin:1987jd}) and, eventually, results in a non-local deformed theory in $d=3$ (as already suggested in~\cite{Seibold:2023zkz}).
Let us denote the three-dimensional coordinates by $(x^0,x^1,y)$, where  $y$ is the compact coordinate. We will often write $(x,y)$ for brevity.
The real scalar field $\Phi(x,y)$ is $2\pi R$-periodic along~$y$ and hence it admits the Fourier expansion 
\begin{equation}\label{eq:KKexpansion}
  \Phi(x,y)
  =\frac{1}{\sqrt{2\pi R}}
  \sum_{n\in\mathbb{Z}} \Phi_n(x)\,e^{i n y /R}\,,\qquad
  \Phi_n(x)^*=\Phi_{-n}(x)\,.
\end{equation}
Defining the $3d$ Lagrangian density of a free massless scalar
\begin{equation}\label{3d0}
    \mathcal{L}_0(\Phi)=-\frac{1}{2}\partial_\mu \Phi\,\partial^\mu \Phi\,,
\end{equation}
upon compactification, we have
\begin{equation}
\mathcal{S}_0=\int \d^2x \int\limits_0^{2\pi R} \d y\,\mathcal{L}_0(\Phi)= \int \mathrm{d}^2\sigma\,\mathcal{L}_{0}(\Phi_n;R)\,,
\end{equation}
where 
\begin{equation}
    \mathcal{L}_{0}(\Phi_n;R)= - \frac{1}{2}
  \sum_{n\in\mathbb{Z}}
  \left(
  \partial_\alpha \Phi_n\,\partial^\alpha \Phi_{-n}
  +  \frac{n^2}{R^2}\, \Phi_n \Phi_{-n}
  \right)\,,
\label{KKmodes}
\end{equation}
which is the Lagrangian for infinitely many free complex scalars of masses $M=n/R$ in the $d=2$ theory. 
Using the results of ~\cite{Bonelli:2018kik,Frolov:2019nrr}, we get the following $T\overline{T}$-deformed Lagrangian density 
\begin{equation}\label{actionKK}
\mathcal{L}_\lambda(\Phi_n;R)
=
\frac{1}{\lambda \mathbf{A}}
\left[
\sqrt{
\lambda \mathbf{A}
\left(
\frac{\lambda}{2}\mathbf{A}\,{\bf P}
-
\sum_{n\in\mathbb{Z}}
\partial_\alpha \Phi_n\,\partial^\alpha \Phi_{-n}
\right)
+1
}
+\,\lambda {\bf V}-1
\right]\,,
\end{equation}
where $\lambda$ denotes the deformation parameter, and
\begin{align}\label{V}
{\bf P}
&=
\sum_{n,m\in\mathbb{Z}}
\Big(
\partial_\alpha \Phi_n\,\partial^\alpha \Phi_{-n}\,
\partial_\beta \Phi_m\,\partial^\beta \Phi_{-m}
-
\partial_\alpha \Phi_n\,\partial^\beta \Phi_{-n}\,
\partial_\beta \Phi_m\,\partial^\alpha \Phi_{-m}
\Big)\,, \notag\\
{\bf V}
&=
\frac12\sum_{n\in\mathbb{Z}}\frac{n^2}{R^2}\,\Phi_n \Phi_{-n}\,,\qquad \mathbf{A}=\frac{\lambda \mathbf{V}}{2}-1\,.
\end{align}
We now want to find a $d=3$, $\lambda$-dependent action whose Kaluza-Klein compactification gives precisely~\eqref{actionKK}. It is easiest to work perturbatively in $\lambda$,
\begin{equation}\label{eq:Lseries}
  \mathcal{L}_{\lambda}(\Phi_n;R)
  = \mathcal{L}_{0}(\Phi_n;R)
    + \frac{\lambda}{2}\,\mathcal{O}_1(\Phi_n;R)
    + O(\lambda^2)\,,\qquad
    \mathcal{O}_1(\Phi_n;R)=\text{det} T_{\alpha\beta}\Big|_{\lambda=0}\,,
\end{equation}
where $\mathcal{O}_1(\Phi_n;R)$ is the deforming operator at the leading order~\cite{Seibold:2023zkz}
\begin{align}\label{eq:L1KK}
  \mathcal{O}_1(\Phi_n;R)
  = -\frac{1}{4}
    &\sum_{n,m\in\mathbb{Z}}\Big(
    \partial_\alpha \Phi_n \,\partial^\alpha \Phi_{-n}\,
    \partial_\beta \Phi_m \,\partial^\beta \Phi_{-m}
    - 2\, 
    \partial_\alpha \Phi_n \,\partial^\beta \Phi_{-n}\,
    \partial_\beta \Phi_m \,\partial^\alpha \Phi_{-m} \notag\\
  &\hspace{4em}
    +\, \frac{m^2 n^2}{R^4} \,
    \Phi_n \Phi_{-n}\, \Phi_m \Phi_{-m}
  \Big)\,.
\end{align}
\paragraph{Leading-order uplift to three dimensions.}
As pointed out in~\cite{Seibold:2023zkz}, the linear term $\mathcal{O}(\Phi_n;R)$ in the deformation cannot be generated by any local operator in three dimensions. Rather, it admits a bilocal realisation along the compact direction. Accordingly, at leading order, one may write
\begin{align}
\mathcal{S}_\lambda
&=\int \mathrm{d}^2x\,
\Big[
\mathcal{L}_{0}(\Phi_n;R)
+\frac{\lambda}{2}\,\mathcal{O}_1(\Phi_n;R)
+O(\lambda^2)
\Big]\notag\\
&=\int \mathrm{d}^2x\int\limits_{0}^{2\pi R}\!\mathrm{d}y\,
\Bigg[
-\frac{1}{2}\,\partial^\mu\Phi\,\partial_\mu\Phi
+\frac{\lambda}{2}\int\limits_{0}^{2\pi R}\!\mathrm{d}y'\,
\mathcal{O}_{1}(x,y;y')
+O(\lambda^2)
\Bigg]\,,
\end{align}
where the bilocal kernel is
\begin{align}\label{eq:nonlocalkernel}
\mathcal{O}_{1}(x,y;y')
=&-\frac{1}{4}\,
\partial_\alpha\Phi(x,y)\,\partial^\alpha\Phi(x,y)\,
\partial_\beta\Phi(x,y')\,\partial^\beta\Phi(x,y')\notag\\
&+\frac{1}{2}\,
\partial_\alpha\Phi(x,y)\,\partial^\alpha\Phi(x,y')\,
\partial_\beta\Phi(x,y')\,\partial^\beta\Phi(x,y)\notag\\
&-\frac{1}{4}\,
\partial_y\Phi(x,y)\,\partial^y\Phi(x,y)\,
\partial_{y'}\Phi(x,y')\,\partial^{y'}\Phi(x,y')\,.
\end{align}
By construction, the integrand is symmetric under exchange of the compact coordinates, $y\longleftrightarrow y'$, so that the $y$ and $y'$ integrations are freely interchangeable.
Let us observe that the kernel~\eqref{eq:nonlocalkernel} can be rewritten compactly as
\begin{align}\label{LO}
\mathcal{O}_1(x,y;y')
&=
\frac12\,
t_{\alpha\beta}(x,y)\,
t^{\alpha\beta}(x,y')
-
\mathcal{J}(x,y)\,\mathcal{J}(x,y')\notag\\
&=\frac12\,\varepsilon^{\alpha\gamma}\varepsilon^{\beta\delta}t_{\alpha\beta}(x,y)t_{\gamma\delta}(x,y')-
\mathcal{J}(x,y)\,\mathcal{J}(x,y')\,.
\end{align}
where
\be
t_{\alpha\beta}(x,y)
\equiv
\partial_\alpha \Phi(x,y)\,\partial_\beta \Phi(x,y)
-\frac{1}{2}\eta_{\alpha\beta}\,\partial_\gamma \Phi(x,y)\,\partial^\gamma \Phi(x,y)\,,
\ee
and
\be
\mathcal{J}(x,y)
\equiv
\frac12\,\partial_y \Phi(x,y)\,\partial^y \Phi(x,y)\,.
\ee
Note that $\mathcal{J}$ is the two-dimensional trace of the stress-energy tensor of the seed theory~\eqref{3d0}, $\mathcal{J}={-\frac 12 }T_{\alpha}{}^{\alpha}$, summed over $\alpha=0,1$, and  $t_{\alpha\beta}$ is the traceless part of $T_{\alpha\beta}$, namely, $t_{\alpha\beta}=T_{\alpha\beta}-\tfrac{1}{2}\eta_{\alpha\beta}\,{T_\gamma{}^{\gamma}}$.
The structure \eqref{LO} is similar to the two-dimensional
$T\overline T$ operator, but is promoted to a point-split object along the
compact direction.\footnote{In full generality, one may also allow for mixed terms coupling $\mathcal{J}$ to the $\tr\, T$. However, these vanish identically in the traceless sector considered here.}

\paragraph{Uplift to three dimensions at higher orders.}\label{paragraph}
Beyond leading order in the deformation parameter, the uplifted operator in three dimensions becomes progressively non-local along the compact direction. Correspondingly, the deformed action admits the schematic expansion 
\begin{equation}\label{int2:3}
\mathcal{S}_\lambda
= \int \mathrm{d}^2x \!\!\int\limits_{0}^{2\pi R}\!\! \mathrm{d}y
\Bigg(\mathcal{L}_0(x,y)+\frac{\lambda}{2}\!\!\int\limits_{0}^{2\pi R}\!\!\d y'\Bigg(\mathcal{O}_1(x,y;y')+\frac{\lambda}{2}\!\!\int\limits_{0}^{2\pi R}\!\!\d y''\Big(\mathcal{O}_2(x,y;y',y'')+\dots\Big)\Bigg)\Bigg),
\end{equation}
where the seed Lagrangian is \eqref{3d0}, and each subsequent term involves one more integration. The first few higher-order kernels can be written in the following form
\begin{equation}\label{NLO}
\mathcal{O}_{2}(x,y;y',y'')=
\mathcal{O}_{1}(x,y;y')\,\mathcal{J}(x,y'')
+\varepsilon^{\alpha\gamma}\varepsilon^{\beta\delta}\,
t_{\alpha\beta}(x,y)\,t_{\gamma\delta}(x,y')\,
\mathcal{L}_0\!\left[\Phi(x,y'')\right]\,,
\end{equation}
and
\begin{align}\label{NNLO}
\mathcal{O}_{3}(x,y;y',y'',y''')
&= \mathcal{O}_{2}(x,y;y',y'')\,\mathcal{J}(x,y''')
+\frac{1}{2}\,\varepsilon^{\alpha\gamma}\varepsilon^{\beta\delta}\,
t_{\alpha\beta}(x,y)\,t_{\gamma\delta}(x,y') \notag\\
&\quad\times
\Bigg[
\mathcal{O}_{1}(x, y'';y''')+
\left(2\,\mathcal{L}_0\!\left[\Phi(x,y'')\right]+\mathcal{J}(x,y'')\right)
2\mathcal{L}_0\!\left[\Phi(x,y''')\right]
\Bigg]\,.
\end{align}
Even in this simple instance, we could not find a natural pattern for the higher-order corrections, though they can, in principle, be worked out order by order. This, together with the non-local nature of the resulting theory, makes the physical interpretation of this deformed model less transparent than we might have hoped.
\section{Dirac--Nambu--Goto--type theories in \texorpdfstring{$d$}{d} dimensions}\label{DNG} 
In  $d=2$ the lightcone gauge-fixed string action yields a $T\overline{T}$ deformation~\cite{Baggio:2018gct,Frolov:2019nrr,Sfondrini:2019smd}. In the simplest case of flat spacetime, this can also be interpreted as the static gauge of the Nambu--Goto (NG) action, up to subtracting ``by hand'' a divergent constant term from the action (more properly, this term arises from the B-field in the lightcone gauge).
We want to consider now the case of a $p$-brane moving in $D$-dimensions, whose action before gauge fixing takes the form~\cite{Dirac:1962iy, Goto:1971ce, Nambu:1974zg}
\begin{equation}
\label{eq:pbraneaction}
    \mathcal S=
    -\frac{1}{\lambda}\int \mathrm{d}^{d}x\sqrt{-\det\Big(\partial_\mu X^A\partial_\nu X^B \,G_{AB}(X)\Big)}\,,
\end{equation}
where $d=p+1$ denotes the world-volume of the brane, $\lambda$ is the inverse of the tension, $A,B=0,1,2,\dots, D-1$ and $G_{AB}(X)$ is the target-space metric tensor. 

Starting from the action~\eqref{eq:pbraneaction} we can obtain a $T\overline{T}$-like flow for $N=D-d$ scalar fields $\Phi^I(x)$, with $I=1,\ldots, N$.  For simplicity, let us consider the case where the first $d$ coordinates describe $d$-dimensional Minkowski space. In this case, the action can be rewritten in the static gauge by setting
\begin{equation}
    X^A = (x^0, x^{1},\dots, x^{d-1}, \sqrt{\lambda}\,\Phi^1,\dots \sqrt{\lambda}\,\Phi^N)\,,\qquad N=D-d\,.
\end{equation}
The rescaling by $\sqrt{\lambda}$, due to which the fields $\Phi^I$ have the canonical dimension of mass $m^{{\frac{d-2}2}}$,  will be important in what follows. 
With this choice, the action~\eqref{eq:pbraneaction} takes the form%
\begin{equation}\label{eq:pbranegaugefixed}
\mathcal S
=
-\frac{1}{\lambda}
\int \mathrm{d}^{d}x\,
\sqrt{
-\det\!\left(
\eta_{\mu\nu}
+
\lambda\, \partial_\mu \Phi^I \partial_\nu \Phi^J G_{IJ}(\Phi)
\right)
}\,,
\end{equation}
where we will assume that $G_{IJ}(\Phi)$ is independent of $\lambda$.

The actions \eqref{eq:pbraneaction} and \eqref{eq:pbranegaugefixed} (like the string in static gauge) are singular in the $\lambda\to0$ limit. For the string ($d=2$), this can be avoided by using lightcone gauge rather than static gauge, or by adding by hand the contribution of a constant B-field. For the $d$-dimensional action, let us just add a constant (`cosmological') term, and set
\begin{equation}\label{eq:braneactionnoV}
\mathcal S_\lambda
=
\int \mathrm{d}^{d}x\, \mathcal L_\lambda
=
\int \mathrm{d}^{d}x
\left(
\frac{\sqrt{-g}}{\lambda}
-
\frac{1}{\lambda}
\sqrt{
-\det\!\left(
g_{\mu\nu}(x)
+
\lambda\, \partial_\mu \Phi^I \partial_\nu \Phi^J G_{IJ}(\Phi)
\right)
}
\right)\, .
\end{equation}
Note that we have also slightly generalised~\eqref{eq:pbranegaugefixed} by allowing for a generic metric $g_{\mu\nu}(x)$  on the $d$-dimensional world-volume (which, for instance, may account for more general gauge-fixing).  The metric $g_{\mu\nu}$ and its inverse $g^{\mu\nu}$ will be used to lower and raise $d$-dimensional worldvolume indices.

In the limit $\lambda \to 0$ the action \eqref{eq:braneactionnoV} reduces to a sigma-model action \cite{Gursey:1959yy, GellMann1960TheAV}
\be\label{eq:flowICnoV}
\mathcal S_{0}=-\frac 12 \int \d^{d}x\,\sqrt{-g}\,g^{\mu\nu}\partial_\mu \Phi^I\partial_\nu \Phi^J G_{IJ}(\Phi)\,.
\ee
Hence, we would like to interpret $\mathcal{S}_\lambda$ in equation~\eqref{eq:braneactionnoV} as a $\lambda$-flow with initial condition $\mathcal{S}_{0}$ given by~\eqref{eq:flowICnoV}, driven by a suitable operator. We want this operator to depend only on the stress-energy tensor and possibly explicitly on the parameter $\lambda$.

\subsection{Flow equation}
We will now show that the theory described by the action \eqref{eq:braneactionnoV} can be viewed as a stress-tensor deformation of the action~\eqref{eq:flowICnoV}, such that the Lagrangian density of \eqref{eq:braneactionnoV} satisfies a flow equation of the type
\be\label{gflow}
\frac{1}{\sqrt{-g}}\frac{\partial\mathcal{L}_\lambda}{\partial\lambda}=\mathcal O\left(T^{\mu\nu},\lambda\right)\,,\qquad
T^{\mu\nu}=\frac 2{\sqrt{-g}}\frac{\partial \mathcal{L}_\lambda}{\partial g_{\mu\nu}},
\ee
under the initial conditions $\mathcal{L}_\lambda|_{\lambda=0}=\mathcal L_{0}$.
Note that $T^{\mu\nu}$ also depends on $\lambda$, though we keep this dependence implicit to avoid burdening our notation.
An explicit form of the stress-tensor flow equation \eqref{gflow} satisfied by the action \eqref{eq:braneactionnoV}
was originally found in~\cite{Blair:2024aqz}. We revisit its derivation here
as part of a general framework that will be further exploited and generalized in the following sections.

To work out the differential equation, it is convenient to introduce the symmetric ``metric'' tensor $\mathcal{G}_{\mu\nu}$ and  its inverse $\mathcal{G}^{\mu\nu}$,  given by%
\footnote{To avoid a confusion, let us note that $\mathcal G^{\mu\nu}\not = g^{\mu\rho}g^{\nu\sigma} \mathcal G_{\rho\sigma}$. The tensor $g^{\mu\rho}g^{\nu\sigma} \mathcal G_{\rho\sigma}$ is not inverse of $\mathcal G_{\mu\nu}$ (i.e., $g^{\mu\rho} \mathcal G_{\rho\sigma}g^{\sigma\nu}\mathcal G_{\nu\kappa}\not=\delta^\mu_\kappa$) and  will never appear in our formulas.}
\be\label{im}
\mathcal{G}_{\mu\nu}\equiv g_{\mu\nu}+\lambda \partial_\mu \Phi^I\partial_\nu \Phi^JG_{IJ}\,,  \qquad \mathcal G^{\mu\rho}\mathcal G_{\rho\nu}=\delta^\mu_\nu\,,\qquad 
\mathcal G\equiv\det \mathcal G_{\mu\nu}\,.
\ee
Using these definitions we find the following form of the $\lambda$-derivative of the Lagrangian density ~\eqref{eq:braneactionnoV} 
\bea\label{flowg}
\frac 1{\sqrt{-g}}\frac{\partial\mathcal L_\lambda}{\partial\lambda}
&=&-\frac 1{\lambda^2}+\frac 1{\lambda^2}\sqrt{ \mathcal G g^{-1}}-\frac 1{2\lambda}\sqrt{\mathcal G g^{-1}}\,\Big(\mathcal G^{\mu\nu}\partial_\mu \Phi^I\partial_\nu \Phi^JG_{IJ}\Big)\nonumber\\
&=&-\frac 1{\lambda^2}+\frac {2-d}{2\lambda^2}\sqrt{ \mathcal G g^{-1}}+\frac 1{2\lambda^2}\sqrt{\mathcal G g^{-1}}\,\mathcal G^{\mu\nu}g_{\mu\nu}\,,
\eea
and furthermore, we get
\begin{equation}\label{def_of_T}
T^{\mu\nu}=\frac 1\lambda\Big(g^{\mu\nu}-\sqrt{ \mathcal G g^{-1}}\,\mathcal G^{\mu\nu}\Big)\,.
\end{equation}
From \eqref{def_of_T}, we construct the following invariants of the stress-energy tensor
\begin{subequations}\label{EMT_in_gG}
    \begin{align}
        \tr\, T = T^{\mu\nu}g_{\mu\nu} &= \frac 1\lambda\Big(d-\sqrt{ \mathcal G g^{-1}}\,\mathcal G^{\mu\nu}g_{\mu\nu}\Big)\,, \label{trTg}
        \\
        T_{\mu\nu}T^{\mu\nu}
&=\frac 1{\lambda^2}\left(-d + 2\lambda \tr\, T+\mathcal G g^{-1}\left(\mathcal G^{\mu\nu}\mathcal G^{\rho\sigma}g_{\nu\rho}g_{\mu\sigma}\right)\right)\,, \label{TTg}
        \\
        \det\left(\delta_\mu^\nu-\lambda T_{\mu}{}^{\nu}\right)&=\Big(\mathcal Gg^{-1} \Big)^{\frac {d-2}2}\,. \label{det(T-gamma)}
    \end{align}
\end{subequations}
Using the relations \eqref{trTg} and \eqref{det(T-gamma)} we bring the equation \eqref{flowg} to the following form
\be\label{eq:flowEqnoV}
\frac 1{\sqrt{-g}}\frac{\partial \mathcal L_\lambda}{\partial \lambda}
=-\frac 1{2\lambda}\left(\tr\, T+\frac{2-d}\lambda\right)+\frac{2-d}{2\lambda^2}\left(\det\left(\delta_\mu^\nu-\lambda T_{\mu}{}^{\nu}\right)\right)^{\frac 1{d-2}}\, .
\ee
We thus reproduce the flow equation of \cite{Blair:2024aqz} governing the $T\overline T$-like deformation of the $d$-dimensional sigma-model action~\eqref{eq:flowICnoV} for $N$ scalar fields, whose solution yields the action~\eqref{eq:braneactionnoV}. 

We would like to treat the flow equation \eqref{eq:flowEqnoV} more generally, i.e., to use it to deform theories which are different from \eqref{eq:flowICnoV}. For this to be possible, one must set initial conditions for the flow at $\lambda=0$ in terms of an action $\mathcal{S}_0$ and of the corresponding stress-energy tensor $T^{\mu\nu}_0$.  Expanding~\eqref{eq:flowEqnoV} at small $\lambda$ and using the well-known decomposition of the determinant over powers of the trace (see e.g.~\cite{muir2003treatise}), we find
\begin{equation}
\label{eq:expandedflow}
    \frac 1{\sqrt{-g}}\frac{\partial \mathcal L_\lambda}{\partial \lambda}=\frac{1}{4}\left[\frac{1}{2-d}(\tr\, T_{0})^2+\tr\left(T_{0}^2\right)\right]+O(\lambda)\,,
\end{equation}
from which we see that the expansion is regular in $d\neq2$ for any seed theory, but it is only well defined in $d=2$ if $\tr\, T_{0}=0$. In other words, in $d\neq 2$ the flow equation~\eqref{eq:flowEqnoV} can be solved perturbatively for any seed theory, but in $d=2$ it only makes sense if the seed theory is conformal. We will return to this point in section~\ref{sec:braneexamples}.

The fact that we can rewrite the equation \eqref{flowg} in terms of $T^{\mu\nu}$ only is not surprising, because
$\mathcal{G}_{\mu\nu}$ and $T_{\mu\nu}$ are both symmetric matrices and hence have the same number of independent invariants which can always be related to each other, like in equations \eqref{trTg}--\eqref{det(T-gamma)}.\footnote{Here and in what follows by ``independent invariants'' of $d\times d$ matrices $\mathcal{G}_{\mu\nu}$ and $T_{\mu\nu}$ we mean functionally independent scalars constructible, respectively, from components of $\mathcal{G}_{\mu\nu}$ or $T_{\mu\nu}$, which are invariant under $SO(1,d-1)$ in the flat case ($g_{\mu\nu}=\eta_{\mu\nu}$). For a general analysis of the structure of invariants of tensor fields of various groups in various dimensions, see e.g. \cite{Cederwall:2025ywy,Elamaran:2025jla}.} We will show, however, that the same is possible for more general actions, which include a scalar field potential, as discussed in the next section, and for certain DBI theories (section \ref{split}).

\subsection{Including a potential for the scalars}\label{pot}
Let us now get the solution of the flow equation \eqref{eq:flowEqnoV} for a seed action
\be\label{eq:flowICnoVV}
\mathcal S_{0}=- \int \d^{d}x\,\sqrt{-g}\,\left(\frac 12g^{\mu\nu}\partial_\mu \Phi^I\partial_\nu \Phi^J G_{IJ}(\Phi)+V(\Phi)\right)\,,
\ee
which includes a potential~$V(\Phi)$. 
Rather than presenting the solution outright, it is instructive to derive it. Let us take a general ansatz
\be\label{pbraneactV}
\mathcal S_\lambda=\int \d^{d}x\left[\frac {\sqrt{-g}}\lambda f(\lambda V)-\frac {\phi(\lambda V)}\lambda\sqrt{-\det\Big(g_{\mu\nu}(x)+\lambda\,\kappa(\lambda V)\, \partial_\mu \Phi^I\partial_\nu \Phi^JG_{IJ}(\Phi)\Big)}\right]\,,
\ee
motivated by the case $V(\Phi)=0$ and by dimensional considerations.
Here $f(z)$, $\phi(z)$ and $\kappa(z)$ are real functions such that 
\begin{equation}\label{conditions}
   f(0)=\phi(0)=\kappa(0)=1\,,\qquad  \phi'(0) - f'(0) = 1,  
\end{equation}
where prime denotes the derivative with respect to the argument $z=\lambda V$.
These initial conditions ensure that when $V=0$, the above action reduces to \eqref{eq:braneactionnoV}, while in the limit $\lambda\to 0$ equation \eqref{pbraneactV} reduces to \eqref{eq:flowICnoVV}.

It is convenient to amend our definition of $\mathcal{G}_{\mu\nu}$ so that
\be\label{kappaG}
\mathcal G_{\mu\nu}=g_{\mu\nu}+\lambda\kappa(\lambda V) \partial_\mu \Phi^I\partial_\nu \Phi^JG_{IJ}(\Phi)\,,
\ee
and similarly for its inverse tensor $\mathcal{G}^{\mu\nu}$ and determinant $\mathcal{G}$.
The $\lambda$-derivative of the Lagrangian density \eqref{pbraneactV} is
\begin{align}\label{flowgV}
        &\frac 1{\sqrt{-g}}\frac{\partial\mathcal L_\lambda}{\partial\lambda}
=-\frac {f-\lambda V f'}{\lambda^2}+\frac {\phi-\lambda V \phi'}{\lambda^2}\sqrt{ \mathcal G g^{-1}}-\frac {\phi(\kappa+\lambda V\kappa')}{2\lambda}\sqrt{\mathcal G g^{-1}}\,\Big(\mathcal G^{\mu\nu}\partial_\mu \Phi^I\partial_\nu \Phi^JG_{IJ}\Big)\notag\\
&=-\frac {f-\lambda V f'}{\lambda^2}+\frac {(2-d)\phi-\lambda V( 2\phi'+d\,\phi\,\kappa^{-1}\kappa')}{2\lambda^2}\sqrt{ \mathcal G g^{-1}}+\frac {\phi(1+\lambda V\kappa^{-1}\kappa')}{2\lambda^2}\sqrt{\mathcal G g^{-1}}\,\mathcal G^{\mu\nu}g_{\mu\nu}
\end{align}
and the stress-energy tensor is now given by
\be\label{EMV}
T^{\mu\nu}=\frac 1\lambda\Big(f\,g^{\mu\nu}-\phi\sqrt{ \mathcal G g^{-1}}\,\mathcal G^{\mu\nu}\Big)\,.
\ee
From this, we have
\be\label{trTVdetTV}
    \tr\, T
    =\frac 1\lambda\Big(f\,d-\phi \sqrt{ \mathcal G g^{-1}}\,\mathcal G^{\mu\nu}g_{\mu\nu}\Big),\quad
    \det\Big(\delta^\nu_\mu-\lambda T_{\mu}{}^\nu\Big)=\det \Big(\delta_\mu^\nu(1-f)+\phi\sqrt{ \mathcal G g^{-1}}\, g_{\mu\rho}\mathcal G^{\rho\nu}\Big)\,.
\ee
Substituting these relations into the right-hand side of \eqref{eq:flowEqnoV}, we get
\begin{align}
\frac{1}{\sqrt{-g}}\frac{\partial\mathcal{L}_\lambda}{\partial\lambda}
&= \frac{1}{2\lambda^2}\left(\phi \sqrt{\mathcal G g^{-1}}\,\mathcal G^{\mu\nu}g_{\mu\nu}-2-d(f-1)\right) \notag\\
&\quad +\frac{2-d}{2\lambda^2}\left[\det\Big(\delta_\mu^\nu(1-f)+\phi\sqrt{\mathcal G g^{-1}}\,g_{\mu\rho}\mathcal G^{\rho\nu}\Big)\right]^{\frac{1}{d-2}}\label{TTflowV}\, .
\end{align}
The requirement that the right-hand side of \eqref{TTflowV} coincides with \eqref{flowgV} 
uniquely determines  the functions $f$, $\phi$ and $\kappa$.
When $d\neq2$ we have
\be\label{fkphig}
\begin{array}{ccc}
     f = 1 \,,\qquad & \kappa= 1\, , \qquad  & 2\lambda V\cdot\phi' (\lambda V) = (2-d)\cdot\phi(\lambda V) \left(1-\phi(\lambda V)^{\frac 2{d-2}}\right) \, .
\end{array}
\ee
The relation for $\phi$ is an ordinary differential equation whose general solution is
\be
\phi (z) = (1-c\, z)^{-\frac{d-2}{2}}\,,
\ee
The constant $c$ is determined by the initial condition $\phi'(0) - f'(0) = 1$, which gives $c=2/(d-2)$.  All in all,
\be\label{phi}
\phi(\lambda V)=\Big(1-\frac {2\lambda V}{d-2}\Big)^{-\frac{d-2}2}.
\ee
Finally, plugging equations $f=1$, $\kappa = 1$ and \eqref{phi} into \eqref{pbraneactV} we find the action
\be\label{f=k=1}
\mathcal S_\lambda=\int \d^{d}x\,\frac 1\lambda\left( {\sqrt{-g}} -{\left(1-\frac {2\lambda V(\Phi)}{d-2}\right)^{-\frac{d-2}2}}\,\sqrt{-\det\Big(g_{\mu\nu}(x)+\lambda 
\partial_\mu \Phi^I\partial_\nu \Phi^JG_{IJ}(\Phi)\Big)}\right)\,,
\ee
valid for $d\neq2$. To the best of our knowledge, this result is new. The case of $d=2$ is special and will be considered in section~\ref{sec:braneexamples}.

\subsection{Single scalar field}\label{single}
The case of a single scalar~$\Phi$ (that is, $N=1$) has already been studied in~\cite{Ferko:2023sps}. It was found therein that a seemingly different differential equation governs the flow of the action~\eqref{eq:braneactionnoV}, namely 
\be\label{Rgflow}
\frac 1{\sqrt{-g}}\,\frac{\partial\mathcal L_\lambda}{\partial\lambda}=\frac 1{2d}T_{\mu\nu}T^{\mu\nu}-\frac 1{d^2}(\tr\, T)^2+\frac{d-2}{d^2}R\,\tr\, T\,,
\ee
where
\be\label{Rg}
R=\sqrt{\frac d{4(d-1)}\Big(T_{\mu\nu}T^{\mu\nu}-\frac 1d (\tr\, T)^2\Big)}\,.
\ee
The same ``Root-$T\overline T$" operator governs stress-tensor flows in other theories in various dimensions \cite{Rodriguez:2021tcz,Tempo:2022ndz,Babaei-Aghbolagh:2022uij,Ferko:2022iru,Conti:2022egv,Ferko:2022cix,Borsato:2022tmu,Ferko:2023sps, Ferko:2023wyi,Morone:2024ffm,Ferko:2024zth,Babaei-Aghbolagh:2024hti,Babaei-Aghbolagh:2024uqp,Babaei-Aghbolagh:2025uoz}, like ModMax and its 6-dimensional counterpart \cite{Bandos:2020jsw,Bandos:2020hgy}.

It so happens that with the initial condition
\begin{equation}
    \mathcal{S}_0=- \frac{1}{2}\int \d^{d}x\,\sqrt{-g}\
g^{\mu\nu}\partial_\mu \Phi\partial_\nu \Phi\, G(\Phi)\,,
\end{equation}
\textit{i.e.\ for  a single scalar field without potential}, the flow equation  \eqref{eq:flowEqnoV} and the one \eqref{Rgflow} of~\cite{Ferko:2023sps}
 give rise to the very same flow, since the operators in \eqref{eq:flowEqnoV} and \eqref{Rgflow} are equivalent along the flow \cite{Blair:2024aqz}.
This is because  $\partial_\mu\Phi\partial^\nu\Phi \,G(\Phi)$ is a degenerate $d\times d$ matrix of rank 1. This can be checked explicitly as further discussed in section \ref{BIexamplesMoreThan2d}, but here it is perhaps more instructive to construct the solution of the flow equation~\eqref{Rgflow} in the case with a potential,
\begin{equation}
\label{eq:S0Nequals1potential}
    \mathcal{S}_0=- \int \d^{d}x\,\sqrt{-g}\ \left(\frac{1}{2}
g^{\mu\nu}\partial_\mu \Phi\partial_\nu \Phi\, G(\Phi)+V(\Phi)\right)\,,
\end{equation}
and show that it coincides with our~\eqref{f=k=1} only in the case $V(\Phi)=0$. Let us do this.

Once again, let us take the ansatz \eqref{pbraneactV}, whose derivative with respect to the parameter $\lambda$ was given in \eqref{flowgV}. 
In the case of a single scalar, it has the form 
\be\label{flowgV1}
\frac 1{\sqrt{-g}}\frac{\partial\mathcal L}{\partial\lambda}
=-\frac {f-\lambda V f'}{\lambda^2}+\frac {\phi-\lambda V( 2\phi'+\phi\kappa^{-1}\kappa')}{2\lambda^2}\sqrt{ \mathcal G g^{-1}}+\frac {\phi(1+\lambda V\kappa^{-1}\kappa')}{2\lambda^2}\sqrt{\mathcal G^{-1} g}\,.
\ee
To get \eqref{flowgV1}, we used the following relations, which are valid for a single scalar matrix \eqref{kappaG}
\begin{align}\label{1sG}
&\mathcal Gg^{-1}=1+\lambda\kappa \partial_\mu\Phi\partial^\mu\Phi G(\Phi)\,, \qquad\mathcal G^{\mu\nu}=g^{\mu\nu}-\lambda\kappa\partial^\mu\Phi\partial^\nu\Phi G(\Phi)\, g\mathcal G^{-1}\,,
\\
&\mathcal G^{\mu\nu}g_{\mu\nu}=d-1 + \mathcal G^{-1} g\,, \qquad \mathcal{G}^{\mu\nu}\mathcal{G}^{\rho\sigma}g_{\nu\rho}g_{\sigma\mu}=d-1+(\mathcal G^{-1} g)^2\,.
\end{align}
Let us now compute the right-hand side of the flow equation \eqref{Rgflow} and match the expression with \eqref{flowgV1}. 
To this end, we reuse the expressions \eqref{EMV} and \eqref{trTVdetTV} for the stress-energy tensor and its trace.
We thus find
\begin{subequations}\label{TT_trT^2_RV}
\begin{align}
T^{\mu\nu}T_{\mu\nu}&=\frac 1{\lambda^{2}}\Big(d\, f^2 -2f\phi\sqrt{ \mathcal G g^{-1}}(d-1+ \mathcal G^{-1}g)+\phi^2 \mathcal{G} g^{-1}\mathcal{G}^{\mu\nu}g_{\mu\rho}\mathcal{G}^{\rho\sigma}g_{\nu\sigma}\Big)\,,\label{TT}\\
(\tr\, T)^2&=\frac 1{\lambda^2}\Big(d^2f^2 -2d\,f\phi \sqrt{ \mathcal G g^{-1}}(d-1+ \mathcal{G}^{-1}g)+\phi^2 \mathcal{G} g^{-1}(d-1+ \mathcal{G}^{-1}g)^2\Big)\,,\label{trT^2}\\
R&=\pm \frac 1 {2\lambda}\sqrt{\mathcal G^{-1} g}\Big\vert\phi(\mathcal Gg^{-1}-1)\Big\vert\label{RV}=\pm \frac 1 {2\lambda}\sqrt{\mathcal G^{-1} g}\, |\lambda\kappa\phi\partial_\mu \Phi\,\partial^\mu\Phi\, G(\Phi)|\,.
\end{align}
\end{subequations}
Therefore, the choice of the $\pm$ sign is correlated with the positive or negative definiteness of \linebreak $\lambda\kappa\phi \partial_\mu \Phi\,\partial^\mu \Phi\, G(\Phi)=\pm |\lambda\kappa\phi\partial_\mu \Phi\,\partial^\mu\Phi\, G(\Phi)|$. 
We thus have
\be\label{ROp}
\frac 1{2d}T_{\mu\nu}T^{\mu\nu}-\frac 1{d^2}(\tr\, T)^2+\frac{d-2}{d^2}R\tr\, T
= -\frac{f^2+\phi^2}{2\lambda^2}+\frac{f\phi}{2\lambda^2}\Big(\sqrt{\mathcal G g^{-1}}+\sqrt{\mathcal G^{-1}g}\Big)\,.
\ee
Comparing \eqref{ROp} with the right hand side of \eqref{flowgV1} we find the relations 
\be\label{fkphirel2}
 f(z)=1+z\frac{\kappa'(z)}{\kappa(z)}\,,\qquad f(z)=1-z\, \frac{\phi'(z)}{\phi(z)}\,,\qquad f(z)-z\,f'(z)=\frac{f(z)^2+\phi(z)^2}2\,
\ee
for $f(z)$, $\phi(z)$ and $\kappa(z)$.
Together with the initial conditions $\phi(0)=\kappa(0)=1$, the first and the second equation imply that $\phi(z)=1/\kappa(z)$.
The third equation in~\eqref{fkphirel2} yields
\be
2 z^2\, \kappa (z) \kappa ''(z)-\left(\kappa (z)-z\, \kappa '(z)\right)^2+1=0\,,
\ee
with initial condition $\kappa(0)=1$. Assuming that $\kappa(z)=1+\mathcal{O}(z)$ has a regular small-$z$ expansion, we find the solution
\be
\kappa(z) = 1 + \beta  z\,.
\ee
The initial condition $\phi'(0)-f'(0)=1$, for which $\mathcal L_\lambda|_{\lambda \to 0}$, fixes $\beta=-1/2$ and thus
\be\label{kfphi}
\kappa(z) = 1 -\frac{z}{2}\,,\qquad f(z) = \frac{1 - z}{1 - \frac{z}{2}}\,,\qquad
\phi(z) = \frac{1}{1 - \frac{z}{2}}\,.
\ee
Finally, we find the solution of the differential equation~\eqref{Rgflow} with initial condition~\eqref{eq:S0Nequals1potential} to be
\be\label{STTVg}
\mathcal S_\lambda=\int \d^{d}x\left(\sqrt{-g}\frac{1-\lambda V(\Phi)}{\lambda\Big(1-\frac\lambda 2 V(\Phi)\Big)}-\frac {\sqrt{-\det\Big(g_{\mu\nu}+\lambda\Big(1-\frac\lambda 2 V(\Phi)\Big) \partial_\mu \Phi\partial_\nu \Phi G(\Phi)\Big)}}{\lambda\Big(1-\frac\lambda 2 V(\Phi)\Big)}\right).
\ee
For $d=2$ this action was derived in \cite{Tateo:2017CDD,Bonelli:2018kik}, for $d=3$ in \cite{Ferko:2023sps} and for a generic $d$ in \cite{Babaei-Aghbolagh:2024hti}.
It is easy to see that \eqref{STTVg} reduces to our~\eqref{f=k=1} with $N=1$ only in the case $V(\Phi)=0$, as we had anticipated.

\subsection{Examples}
\label{NGexamples}
Let us now consider how the flow equation \eqref{eq:flowEqnoV} can be expressed in various dimensions.
\paragraph{String ($d=2$).}\label{sec:braneexamples}
In the $d=2$ case the action \eqref{eq:braneactionnoV} is a generalisation of the bosonic string in the lightcone gauge, which can be recovered by setting~$g_{\mu\nu}=\eta_{\mu\nu}$. It is well known that this satisfies the $T\overline{T}$ flow driven by $-\tfrac12\text{det}[T_\mu{}^\nu]$. However, this cannot be seen directly from a $d\to 2$ limit of~\eqref{eq:flowEqnoV} for a generic stress-energy tensor $T^{\mu\nu}$, as the limit is not well defined.  Instead, we restart from equation~\eqref{flowg} for $d=2$ and use~\eqref{trTg} to replace its last term on the right-hand side with $\text{Tr}\,T$, thus getting
\be\label{traceflow}
\frac 1{\sqrt{-g}}\,\frac{\partial\mathcal L_\lambda}{\partial\lambda}=-\frac{1}{2\lambda}\text{Tr}\,T\,.
\ee
Setting $d=2$  in \eqref{det(T-gamma)} we further find that the NG stress-energy tensor satisfies the condition
\be\label{det=tr}
\det(\delta^\nu_\mu-\lambda T_{\mu}{}^{\nu})=1 \qquad \Rightarrow \qquad \text{Tr}\,T=\lambda\det T_{\mu}{}^{\nu},
\ee
using which we recover the conventional $T\overline T$ flow equation
\be\label{ttbar}
\frac 1{\sqrt{-g}}\,\frac{\partial\mathcal L_\lambda}{\partial\lambda}=-\frac{1}{2}\det T_{\mu}{}^{\nu}\,.
\ee
Equation~\eqref{det=tr} is not valid for a generic stress-energy tensor. However, such relations (dubbed ``trace-flow equations'') between $\tr\, T$ and $T\overline T$-type operators $\mathcal O(T,\lambda)$, like \eqref{ttbar}, may appear and have been studied in deformed CFTs in different space-time dimensions from various perspectives.%
\footnote{See for instance~\cite{Osborn:1991gm,Cavaglia:2016oda,McGough:2016lol,Hartman:2018tkw,Kraus:2018xrn,Khoeini-Moghaddam:2020ymm,Babaei-Aghbolagh:2020kjg,Ferko:2023ruw,Ferko:2023sps,Ferko:2023wyi,Nix:2025plr}.}
Of course the $T\overline{T}$ flow equation~\eqref{ttbar} makes sense for any seed theory, while~\eqref{traceflow} produces a trivial flow for any CFT, and it is singular if the seed theory is not conformal --- like in the case of a scalar field with potential.
Recall that the $T\overline{T}$ flow~\eqref{ttbar} for a scalar field with potential is given by~\cite{Tateo:2017CDD, Bonelli:2018kik,Frolov:2019nrr}
\be\label{STTVp}
\mathcal L_\lambda
=\sqrt{-g}\,\frac{1-\lambda V(\Phi)}{\lambda\Big(1-\frac{\lambda}{2}V\Big)}
-\frac{\sqrt{-\det\Big(g_{\mu\nu}+\lambda\left(1-\frac{\lambda}{2}V\right)\,
\partial_\mu \Phi^I\,\partial_\nu \Phi^J\,G_{IJ}(\Phi)\Big)}}{\lambda\Big(1-\frac{\lambda}{2}V\Big)}\,.
\ee
Comparing~\eqref{STTVp} with~\eqref{STTVg}, we see that they display the same functional dependence on the potential $V$. However, in $d=2$ the number of scalar fields $\Phi^I$ can be arbitrary.

\paragraph{Membrane $(d=3)$.}
In this case, the action \eqref{eq:braneactionnoV} can be thought of as the worldvolume theory of a membrane. 
Its flow equation \eqref{eq:flowEqnoV} takes the form
\be\label{membraneflowTT}
\frac 1{\sqrt{-g}}\,\frac{\partial\mathcal L_\lambda}{\partial\lambda}=-\frac 1{2\lambda}\left(\tr\, T-\frac 1\lambda\right)-\frac{1}{2\lambda^2}\,\det \left(\delta_{\mu}^\nu-\lambda T_\mu{}^{\nu}\right)\,.
\ee
In this case  
\bea\label{det(T-I)}
\det \left(\delta_{\mu}^\nu-\lambda T_\mu{}^{\nu}\right)
&=&1-\lambda\tr\, T+\frac{\lambda^2} 2\Big
[(\tr\, T)^2-\tr\left(T^2\right)\Big]-\lambda^3 \det \left(T_{\mu}{}^{\nu}\right)
\eea
and the  flow equation \eqref{membraneflowTT} reduces to
\bea\label{membraneflowTT1}
\frac 1{\sqrt{-g}}\,\frac{\partial\mathcal L_\lambda}{\partial\lambda}&=&\frac{\lambda}{2}\,\det (T_{\mu}{}^{\nu})+\frac 14\Big[\tr\left(T^2\right)-(\tr\, T)^2\Big]\,.
\eea
The same $3d$ flow equation was also derived in~\cite{Tsolakidis:2024wut} from massive gravity.
It is interesting to note \cite{Blair:2024aqz} that the right-hand side features the determinant and, in the square bracket, the combination of traces that yields the determinant of a $2\times2$ matrix.
However, this pattern does not hold for higher $d$.
We also note that for the special case of a single scalar field $\Phi(x)$, the flow equation was found in \cite{Ferko:2023sps} in the following form 
\be\label{ferkoetal}
    \frac 1{\sqrt{-g}}\,\frac{\partial\mathcal L_\lambda}{\partial\lambda}=\frac 16T^{\mu\nu}T_{\mu\nu}-\frac 19 (\tr\, T)^2+\frac 19 R\,\tr\, T
    ~~~\mathrm{with}~~~
    R=\sqrt{\frac 38\left( T^{\mu\nu}T_{\mu\nu}-\frac 13(\tr\, T)^2\right)} \, .
\ee
As in the case of a generic space-time dimension $d$, the flow equation~\eqref{membraneflowTT1} reduces to \eqref{ferkoetal} only for the case of a single scalar field without potential.
If the seed theory has a potential like in \eqref{eq:flowICnoVV}, the solution of our flow equation~\eqref{membraneflowTT1} in $d=3$ is
\begin{equation}
    \mathcal{L}_\lambda
    = \frac{\sqrt{-g}}{\lambda}
      - \frac{1}{\lambda}
        \sqrt{-\frac{\det\Bigl(g_{\mu\nu}
        + \lambda\, \partial_\mu \Phi^I \partial_\nu \Phi^J\, G_{IJ}(\Phi)\Bigr)}{1-2 \lambda V(\Phi)}}\,.
\end{equation}
As we remarked in the previous subsection, this differs from the solution of \eqref{ferkoetal} of~\cite{Ferko:2023sps}, which is instead given by~\eqref{STTVg}.

\paragraph{Three-brane $(d=4)$.}
As previously for $d=3$, this case can be thought of as a stress-tensor deformation describing a four-dimensional worldvolume action of a 3-brane. The flow equation \eqref{eq:flowEqnoV} becomes
\bea\label{flowTT3}
\frac 1{\sqrt{-g}}\frac{\partial\mathcal L_\lambda}{\partial\lambda}
&=& -\frac 1{2\lambda}\left(\tr\, T-\frac 2{\lambda}\right)-\frac 1{\lambda^2}\sqrt{\det\left(\delta_\mu^\nu -\lambda T_{\mu}{}^{\nu}\right)}\,.
\eea
Following~\eqref{eq:expandedflow}, one can expand the square root up to the second order in $\lambda$ and see that the resulting series is regular in the limit $\lambda \to 0$.

In higher dimensions, while the construction remains under control, the deforming (determinant) operator involves roots of increasing order, whose physical interpretation and implications are not yet fully understood (see
\cite{Cardy:2018sdv,Bonelli:2018kik,Hou:2022csf} for similar structures in the literature).

\section{Born--Infeld-type theories}\label{BI}
The case of the $d$-dimensional pure BI theory (i.e., in the absence of the scalar fields) was not analysed in \cite{Blair:2024aqz}, though a particular $d=3$ example can be recovered from their discussion of a $3d$ DBI theory upon setting the scalar fields to zero therein. 

So, the goal of this section is to derive a universal \(T\overline T\)-like flow equation for the $d$-dimensional Born-Infeld actions
\begin{equation}\label{eq:BIgeneralgamma}
    \mathcal S_{\lambda}=
    \int \d^{d}x\,\left(\frac{\sqrt{-g}}{\lambda}-\frac 1\lambda\sqrt{-\det\left(g_{\mu\nu}(x)+\sqrt{\lambda}F_{\mu\nu}\right)}\right)\,.
\end{equation}
where $F_{\mu\nu}=\partial_\mu A_{\nu}(x)-\partial_\nu A_\mu(x)$ is the antisymmetric field strength of the Abelian vector field~$A_\mu(x)$.
At $\lambda\to 0$ we have Maxwell's action as the initial condition
\begin{equation}\label{maxwell}
    \mathcal S_{0}=-\frac{1}{4}\int \d^{d}x\,\sqrt{-g}\,F_{\mu\nu}F^{\mu\nu}\,.
\end{equation}
The Hilbert stress-energy tensor%
\footnote{Using the Hilbert stress-energy tensor ensures that we are dealing with a symmetric tensor.  For the scalar field theories it coincides with the canonical Noether stress tensor, but for the gauge vector field it does not. See~\cite{Baggio:2018rpv} for a discussion of the equivalence of $T\overline{T}$ deformations sourced by apparently different but equivalent stress-energy tensors.}
of the theory is 
\begin{equation}\label{BIEMT}
    T^{\mu\nu}
    =
    \frac{2}{\sqrt{-g}}
    \frac{\partial \mathcal{L}_\lambda}{\partial g_{\mu\nu}}
    =
    \frac{1}{\lambda}
    \left(
        g^{\mu\nu}
        -
        \sqrt{\mathcal{G}g^{-1}}
        \mathcal{G}^{(\mu\nu)}
    \right)\,,
\end{equation}
where, in analogy with the previous section, we introduced 
\be
\mathcal{G}_{\mu\nu} = g_{\mu\nu} + \sqrt{\lambda}\, F_{\mu\nu} \; .
\ee
With $\mathcal G^{(\mu\nu)}$ we denote the symmetric part of the inverse of $\mathcal {G}_{\mu\nu}$. 
The variation of the Lagrangian density  \eqref{eq:BIgeneralgamma} with respect to the parameter $\lambda$ is
\begin{equation}\label{BIflow}
   \frac{1}{\sqrt{-g}}
    \frac {\partial\mathcal{L}_\lambda}{\partial\lambda} =
    -\frac{1}{\lambda^2}
    +
    \frac{4 - d}{4\,\lambda^2} \sqrt{\mathcal Gg^{-1}}
    +
    \frac{1}{4\, \lambda^2}\sqrt{\mathcal Gg^{-1}}\, \mathcal{G}^{\mu\nu} g_{\nu\mu}\,.
\end{equation}
The right-hand side of the above equation can be expressed in terms of the stress-energy tensor with the use of the following relations
\begin{equation}
    \label{GgT}
\frac{1}{4\, \lambda^2}\sqrt{\mathcal{G}g^{-1}}\, \mathcal{G}^{\mu\nu} g_{\mu\nu}
   =
   \frac{d}{4\, \lambda^2} 
   -
   \frac{1}{4\, \lambda}\; \tr\, T\,,\qquad  
    \det \left(\delta_{\mu}^\nu-\lambda T_{\mu}{}^{\nu}\right)
    =\left(\sqrt{\mathcal{G}g^{-1}}\right)^{d-4}\,.
\end{equation}
To derive the second equation of \eqref{GgT}, we used the identity \cite{Bandos:2006wb}
\be\label{BI_Gmatinvsym_relation}
\mathcal G^{(\mu\nu)}=\frac 12 (\mathcal G^{\mu\nu}+\mathcal G^{\nu\mu})=\frac 12 \mathcal G^{\mu\rho}\Big(\delta_{\rho}^\nu+\mathcal G_{\rho\lambda}(\mathcal G^{-\text{T}})^{\lambda\nu}\Big)= \mathcal G^{\mu\rho} \mathcal G_{(\rho\lambda)}(\mathcal G^{-\text{T}})^{\lambda\nu}=\mathcal G^{\mu\rho} g_{\rho\lambda}(\mathcal G^{-\text{T}})^{\lambda\nu}\,,
\ee
where $ (\mathcal G^{-\text{T}})^{\mu\nu}=\mathcal G^{\nu\mu}$. The above identity implies  that 
\begin{equation}\label{detSIM}
    \det \mathcal{G}^{(\mu\nu)}={\mathcal G^{-2}}g~.
\end{equation}
 Thus, we find the stress-tensor flow equation associated with the $d$-dimensional BI-type theory in the following form 
\begin{equation}\label{BIflowT}
\displaystyle   \frac{1}{\sqrt{-g}}
    \frac {\partial\mathcal{L}_\lambda}{\partial\lambda} 
    =
    \frac{d-4}{4\, \lambda^2}
    -
    \frac{\tr\,T}{4 \,\lambda}
    +
    \frac{4-d}{4\, \lambda^2}
    \left(
        \det \left(\delta_\mu^\nu - \lambda T_\mu^{~\nu}\right)
     \right)^{\frac{1}{d-4}} \,.
\end{equation}

Once again, this expression is regular in the $\lambda\to0$ limit.
The expression is ill-defined for a general $T^{\mu\nu}$ as $d\to4$, but it is regular on the stress-energy tensor~\eqref{BIEMT} by virtue of the second equation in~\eqref{GgT}. This is similar to what we saw in the previous section for the brane-like flow in the limit $d\to2$.

It is curious to notice that the BI-type flow equation~\eqref{BIflowT}
in $d$ dimensions has the same algebraic structure as the
DNG flow equation~\eqref{eq:flowEqnoV} in $d-2$ dimensions.
More precisely, the right--hand side of~\eqref{eq:flowEqnoV}, upon the rescalings
$\mathcal{L}\rightarrow \mathcal{L}/ {\sqrt{2}}$ and $
\lambda\rightarrow \sqrt{2}\,\lambda\,$,
coincides algebraically with the right--hand side of~\eqref{BIflowT}.
Next, we consider the flow for particular values of~$d$, and compare with what we found in section~\ref{NGexamples} for DNG flows.

\subsection{Two dimensions}
\label{sec_BI_2d}
Like the $d=2$ Maxwell or Yang--Mills theory \cite{Cordes:1994fc}, the BI theory in $d=2$ does not have dynamical (propagating) degrees of freedom since the equations of motion
\be\nabla_\mu \left( \frac {\partial\mathcal L}{\partial F_{\mu\nu}}\right)=0
\ee
imply that the field strength $F_{\mu\nu}\equiv{\sqrt{-g }}\,\varepsilon_{\mu\nu} f$ is constant ($\partial_\mu f=0$). Still, at the level of the action, it can be viewed as a $T\overline T$-like deformation of Maxwell's theory. For this case, from equation \eqref{BIflowT} we get
\bea\label{BIp=1}
\frac{1}{\sqrt{-g}}
    \frac {\partial\mathcal{L}_\lambda}{\partial\lambda}&=&-\frac 1{2\lambda^2}-\frac 1{4\lambda} \tr\, T+\frac{1}{2\lambda^2} \Big(\det\Big(\delta_\mu^\nu-\lambda T_{\mu}{}^\nu\Big)\Big)^{-\frac 1 2}\,.
\eea
This can be further simplified using the following relations%
\footnote{In our conventions,  $\varepsilon_{\mu\nu}\varepsilon^{\nu\rho} =\delta_\mu^\rho$, since $\varepsilon_{\mu\nu}=-\varepsilon_{\nu\mu},\varepsilon_{01}=-\varepsilon^{01}=1$.}
\be\label{BIinverse}
\mathcal{G}_{\mu\nu}=g_{\mu\nu}+\sqrt{\lambda} F_{\mu\nu}\equiv g_{\mu\nu}+{\sqrt{-g\lambda }}\, \varepsilon_{\mu\nu} f\,, \qquad \mathcal{G}^{\mu\nu}=\frac {g^{\mu\nu}-{\sqrt{-g^{-1}\lambda }} \,\varepsilon^{\mu\nu} f}{1-\lambda f^2}\,.
\ee
Next, we find that 
\be
\mathcal{G}^{(\mu\nu)}=\frac {g^{\mu\nu}}{1-\lambda f^2}\,,\qquad \mathcal{G}=g (1-\lambda f^2)\,,\qquad
T^{\mu\nu}=-g^{\mu\nu}\frac{1-\sqrt{1-\lambda f^2}}{\lambda\sqrt{1-\lambda f^2}} \, .
\ee
Consequently, we establish the equality
\begin{equation}
\label{eq:det-trace-id-BI}
    \det T_\mu^{~\nu}
    =
    \frac{1}{4} (\tr\, T )^2\,,
\end{equation}
valid for this particular stress-energy tensor.
Eventually, \eqref{BIp=1} takes the compact form 
\be\label{eq:twoflows}
\frac{1}{\sqrt{-g}}
    \frac {\partial\mathcal{L}_\lambda}{\partial\lambda}=
  \frac {\det T_{\mu}{}^{\nu}}{2-\lambda \tr\, T}\,,\qquad\text{or}\qquad 
\frac{1}{\sqrt{-g}}
    \frac {\partial\mathcal{L}_\lambda}{\partial\lambda}=
  \frac{1}{4}\,\frac {(\tr\, T )^2}{2-\lambda \tr\, T}\,,
\ee
or any combination thereof which can be obtained by using~\eqref{eq:det-trace-id-BI}.
It is worth noting that the first equation in~\eqref{eq:twoflows} is closely related to the flow used to deform $d=2$ Maxwell and Yang-Mills theories in \cite{Conti:2018jho,Ireland:2019vvj,Brennan:2019azg,Griguolo:2022xcj,Griguolo:2022hek}, but it differs from the latter by the presence of the $\tr\, T$ in its denominator. Indeed, in those cases, the resulting theories were not of a BI type.

\subsection{More than two dimensions}
\label{BIexamplesMoreThan2d}
\paragraph{The case  $d=3$.}
Here the flow equation~\eqref{BIflowT} becomes
\be\label{p=2}
\frac{1}{\sqrt{-g}}
    \frac {\partial\mathcal{L}_\lambda}{\partial\lambda}=-\frac 1{4\lambda^2}-\frac 1{4\lambda} \tr\, T+\frac{1}{4\lambda^2} \Big(\det\Big(\delta_\mu^\nu-\lambda T_{\mu}{}^\nu\Big)\Big)^{-1}\,.
\ee 
Using the results of \cite{Ferko:2023sps}, this flow equation can be rewritten in the following forms
\be\label{RBI}
\frac{1}{\sqrt{-g}}
    \frac {\partial\mathcal{L}_\lambda}{\partial\lambda}=-\frac 1{3\lambda}\Big(\tr\, T-R\Big)=\frac 16T^{\mu\nu}T_{\mu\nu}-\frac 19 (\tr\, T)^2+\frac 19 R\tr\, T\,,
\ee
where $R$ was defined in \eqref{ferkoetal}.

Some what remarkably at the first glance, the deformation operator is identical to that in \eqref{ferkoetal} of the $3d$ DNG one scalar field case, which is the alternative form of \eqref{membraneflowTT}  (for $N=1$). This coincidence can be attributed to the fact that in $d=3$ the massless vector field is dual to a scalar field.
Therefore, since in the $3d$ DNG case and  in the $3d$ BI case the form of the determinants is very similar, namely
\be
\det (g_{\mu\nu}+\lambda\partial_\mu\Phi\partial_\nu\Phi G(\Phi))=g\Big(1+\lambda (\partial_\mu\Phi\partial^\mu\Phi G(\Phi)\Big)
\ee
and 
\be
\det\left(g_{\mu\nu}+\sqrt{\lambda}F_{\mu\nu}\right)=g\left(1+\frac{\lambda}2 F_{\mu\nu}F^{\mu\nu}\right)=g(1-{\lambda}f^\mu f_\mu),
\ee
where $f^\mu=\frac 1{2\sqrt{-g}}\varepsilon^{\mu\nu\rho}F_{\nu\rho}$,
one can expect (and this is indeed the case) that  also in the $3d$ BI case the deformation operator \eqref{RBI} can be rewritten in the form similar to that in \eqref{membraneflowTT}, i.e.
\begin{equation}\label{newbi}
\frac{1}{\sqrt{-g}}
\frac{\partial \mathcal{L}_\lambda}{\partial \lambda}
=
-\frac{1}{2\lambda}
\left(\tr\, T-\frac{1}{\lambda}\right)-\frac{1}{2\lambda^{2}}\,
\det\!\left(\delta_{\mu}^{\nu}-\lambda T_{\mu}{}^{\nu}\right)
\, .
\end{equation}
We thus get three alternative forms \eqref{p=2}, \eqref{RBI}, and \eqref{newbi} of the stress-energy operator which drives the flow of the $3d$ Maxwell action to the BI one.
It is worth noting that \eqref{newbi} takes the same form as the DNG equation~\eqref{membraneflowTT}. This will allow us to combine these two flows and to derive an equation for a $3d$ DBI flow in section~\ref{split}.

\paragraph{The case  $d=4$.}
From the  $d$-dimensional equation \eqref{BIflow}, using \eqref{GgT}, we find the concise four-dimensional version, which is similar to the $2d$ equation \eqref{traceflow}, namely
\be\label{p=3}
\frac{1}{\sqrt{-g}}
    \frac {\partial\mathcal{L}_\lambda}{\partial\lambda}=-\frac 1{4\lambda} \tr\, T\,.
\ee
Similarly to what we did in two dimensions, we can exploit identities valid for the BI stress-energy tensor to recast~\eqref{p=3}  as follows
\be\label{p=31}
\frac{1}{\sqrt{-g}}
    \frac {\partial\mathcal{L}_\lambda}{\partial\lambda}=-\frac 12 \sqrt{\det  T_\mu{}^\nu}= \frac 18\Big(T^{\mu\nu}T_{\mu\nu}-\frac 12 (\tr\, T)^2\Big)\,,
\ee
which are the forms of the operator originally found for the $4d$ BI theory in \cite{Conti:2018jho}. 

Note also that from the second equation in \eqref{GgT} it again follows that 
\be\label{4det=1}
\det (\delta_\mu^\nu - \lambda T_\mu{}^\nu) = 1\,.
\ee
This relation can be used to obtain yet another flow equation
\begin{equation}
\begin{aligned}
\frac{1}{\sqrt{-g}}\frac{\partial\mathcal{L}_\lambda}{\partial\lambda}
&=-
\frac{1}{8}\Big((\tr\, T)^2 - \tr(T^2)\Big)
\\
&\hspace{2.2em}
+\frac{\lambda}{24}\Big(
(\tr\, T)^3
-3\,\tr(T^2)\tr\, T
+2\,\tr\, (T^3)
\Big)
-\frac{\lambda^2}4 \det T_{\mu}{}^{\nu}
\,,
\end{aligned}
\end{equation}
which was first derived in~\cite{Nix:2025plr}.

\paragraph{The case $d=5$.}
In this case,%
\footnote{For potential applications of this setup in cosmology, see e.g \cite{Allahverdizadeh:2013oha}.}
the flow equation~\eqref{BIflowT} takes the form
\be\label{p=4}
\frac{1}{\sqrt{-g}}
    \frac {\partial\mathcal{L}_\lambda}{\partial\lambda}=\frac 1{4\lambda^2}-\frac 1{4\lambda} \tr\, T-\frac{1}{4\lambda^2} \det\Big(\delta_\mu^{\nu}-\lambda T_{\mu}{}^{\nu}\Big)\,.
\ee
It is interesting to write explicitly the form of the above determinant, which terminates at the fifth-order in~$\lambda$, so that the right-hand side as a whole is a cubic polynomial in~$\lambda$. We have
\be\label{p=41}
\begin{split}
\frac{1}{\sqrt{-g}} &
    \frac {\partial\mathcal{L}_\lambda}{\partial\lambda}
    =
    \frac {\lambda^3}4 \det T_{\mu}{}^{\nu}\\
    +&
    \frac 18\Big[\tr (T^2)-(\tr\, T)^2\Big]
    -
    \frac\lambda {24}\Big[(\tr\, T)^3-3(\tr\, T)\tr (T^2)+2\tr(T^3)\Big]\\
    +& 
    \frac{\lambda^2}{96}\Big[(\tr\, T)^4-6(\tr\, T)^2\tr(T^2)+3(\tr(T^2))^2+8(\tr\, T)\tr(T^3)-6\tr (T^4)\Big]\,.
\end{split}
\ee
In higher dimensions, although still under control, the deforming operator involves roots of progressively increasing order, whose physical interpretation remains more elusive.

\section{Dirac--Born--Infeld-type theories}\label{split}
Let us now consider DBI actions, containing both the scalar and gauge fields,
\begin{equation}\label{DBIaction}
    \mathcal S_\lambda
    =\int {\d}^d x\left(
    \frac{\sqrt{-g}}{\lambda}
    -
    \frac{1}{\lambda}
    \sqrt{
    -\det \left(
        g_{\mu\nu}+ \lambda\, \partial_\mu \Phi^I\, \partial_\nu \Phi^J\, G_{IJ}(\Phi) + \sqrt{\lambda}\, F_{\mu\nu} 
    \right)}\right)
    \,.
\end{equation}
 For $F_{\mu\nu}=0$, this reduces to the type of the action considered in section~\ref{DNG}, while for $\Phi=0$ it reduces to those considered in section~\ref{BI}.
In the limit $\lambda\to 0$ the action \eqref{DBIaction} reduces to
\begin{equation}\label{DBI_L0}
   \mathcal{S}_0
= -\frac{1}{2}
\int \mathrm{d}^d x \, \sqrt{-g}\,
\left(
g^{\mu\nu}\partial_\mu \Phi^I\, \partial_\nu \Phi^J\, G_{IJ}(\Phi)
+\frac{1}{2}
F_{\mu\nu}F^{\mu\nu}
\right) \, .
\end{equation}
The matrix 
\be\label{mcGDBI}
\mathcal G_{\mu\nu}= g_{\mu\nu}+ \lambda\, \partial_\mu \Phi^I\, \partial_\nu \Phi^J\, G_{IJ}(\Phi) + \sqrt{\lambda}\, F_{\mu\nu}
\ee
appearing in \eqref{DBIaction} is generic (neither symmetric nor anti-symmetric). Therefore, the number of Lorentz invariants that we may construct from this matrix exceeds the number $d$ of the invariants of the symmetric $d$-dimensional DBI stress-energy tensor, whose form is the same as in equation \eqref{BIEMT}. Hence, in general, it is unclear whether the DBI action may be regarded as stress-tensor deformation of \eqref{DBI_L0}. 
This issue was addressed in \cite{Blair:2024aqz} where it was shown that in $d\neq3$ the  $\lambda$-derivative of the DBI action depends not only on invariants of $T_{\mu\nu}$ but also on those of $F_{\mu\nu}$. The case of the $3d$ DBI turns out to be special. As was shown in \cite{Blair:2024aqz}, its action satisfies the same stress-tensor flow equation as the $3d$ scalar field action \eqref{eq:braneactionnoV} and the $3d$ BI action, namely equation \eqref{newbi}. In this section, we revisit these results.

The $\lambda$-derivative of the Lagrangian density in \eqref{DBIaction} is
\be\label{ldDBI}
\begin{split}
\frac 1{\sqrt{-g}}\frac{\partial\mathcal L_\lambda}{\partial\lambda}
&=
-\frac 1{\lambda^2}+\frac 1{\lambda^2}\sqrt{ \mathcal G g^{-1}}
-
\frac 1{2\lambda}\sqrt{\mathcal G g^{-1}}\,
G^{\mu\nu}\left(\mathcal\partial_\mu \Phi^I\partial_\nu \Phi^JG_{IJ} + \frac{\sqrt{\lambda}}{2} F_{\nu\mu} \right)
\\
&=
-\frac 1{\lambda^2}+\frac {2-d}{2\lambda^2}\sqrt{ \mathcal G g^{-1}}+\frac 1{2\lambda^2}\sqrt{\mathcal G g^{-1}}\,\mathcal G^{\mu\nu}g_{\mu\nu}+\frac 1{4\lambda\sqrt{\lambda}}\sqrt{\mathcal G g^{-1}}\,\Big(\mathcal G^{\mu\nu}F_{\nu\mu}\Big)\,.
\end{split}
\ee
The stress-energy tensor is similar to \eqref{BIEMT}, but with $\mathcal G^{\mu\nu}$ being the inverse of \eqref{mcGDBI}.
The challenge now is to rewrite the right-hand side of \eqref{ldDBI} entirely in terms of invariants of the stress-energy tensor, i.e., to see whether the DBI action can be viewed as a $T\overline T$-like deformation of \eqref{DBI_L0}. As we mentioned, the problem is quite nontrivial compared to the previous cases of the pure scalar and pure gauge field theories, because the number of the invariants of $\mathcal G_{\mu\nu}$ exceeds those of $T_{\mu\nu}$, and also due to the appearance of the additional (last) term in \eqref{ldDBI}. Following \cite{Blair:2024aqz}, we will address this problem for the simplest cases of $d=2$ and $d=3$.

\subsection{Two dimensions}\label{DBI2}
In this case, the equation \eqref{ldDBI} simplifies to
\bea\label{ldDBI2}
\frac 1{\sqrt{-g}}\frac{\partial\mathcal L
_\lambda}{\partial\lambda}&=&
-\frac 1{\lambda^2}+\frac 1{2\lambda^2}\sqrt{\mathcal G g^{-1}}\,\mathcal G^{\mu\nu}g_{\mu\nu}+\frac 1{4\lambda\sqrt{\lambda}}\sqrt{\mathcal G g^{-1}}\,\Big(\mathcal G^{\mu\nu}F_{\nu\mu}\Big)\nonumber\\
&=&-\frac 1{2\lambda}\tr\, T+\frac 1{4\lambda\sqrt{\lambda}}\sqrt{\mathcal G g^{-1}}\,\Big(\mathcal G^{\mu\nu}F_{\nu\mu}\Big)\,.
\eea
To manipulate the last term, it is useful to write the matrix \eqref{mcGDBI}
as 
\be\label{mcGDBIh}
\mathcal G_{\mu\nu}= h_{\mu\nu} + \sqrt{\lambda}\, F_{\mu\nu}\,\qquad{\rm with}\qquad h_{\mu\nu}= g_{\mu\nu}+ \lambda\, \partial_\mu \Phi^I\, \partial_\nu \Phi^J\, G_{IJ}(\Phi)\,.
\ee
Since in $d=2$ the field-strength tensor is proportional to the Levi--Civita tensor, $F_{\mu\nu}=\sqrt{-h}\, \varepsilon_{\mu\nu}f^{(h)}$, it is easy to compute the inverse matrix $\mathcal G^{\mu\nu}$ and other quantities.\footnote{To this end it is useful to keep in mind that  $\varepsilon^{\mu\nu} = - h\cdot h^{\mu\rho} h^{\nu\sigma} \varepsilon_{\rho\sigma}=-g\cdot g^{\mu\rho} g^{\nu\sigma} \varepsilon_{\rho\sigma}$ and $\varepsilon^{\mu\nu} \varepsilon_{\nu\rho} = \delta_\rho^\mu$\,. } We have already done this in section \ref{sec_BI_2d} for $h_{\mu\nu}=g_{\mu\nu}$, cf.~\eqref{BIinverse}. 
In the case under consideration, we have 
\be\label{G-1}
\mathcal G^{\mu\nu}=\frac{h^{\mu\nu}-\sqrt{\lambda}h^{\mu\rho}h^{\nu\sigma}F_{\rho\sigma}}{1+\frac\lambda 2h^{\kappa\lambda}h^{\alpha\beta}F_{\kappa\alpha}F_{\lambda\beta}}=\Big(h^{\mu\nu}-\sqrt{\lambda}h^{\mu\rho}h^{\nu\sigma}F_{\rho\sigma}\Big)\mathcal G^{-1}h\,,
\ee
where $h^{\mu\nu}$ is the inverse of $h_{\mu\nu}$, such that
\be\label{h-1)}
h^{\mu\rho}h_{\rho\nu}=\delta_\nu^\nu\,,\qquad h\equiv\det h_{\mu\nu}\,,\qquad
\mathcal G\equiv\det \mathcal G_{\mu\nu}=h\Big(1+
\frac\lambda 2h^{\kappa\lambda}h^{\alpha\beta}F_{\kappa\alpha}F_{\lambda\beta}\big),
\ee
and hence
\be\label{GF1}
\mathcal G^{\mu\nu}F_{\nu\mu}= \sqrt{\lambda}(h^{\kappa\lambda}h^{\alpha\beta}F_{\kappa\alpha}F_{\lambda\beta})\,\mathcal G^{-1}h=\frac 2{\sqrt{\lambda}}\mathcal (1-\mathcal G^{-1}h)\,.
\ee
Now,  similar to the pure BI case (cf.~\eqref{detSIM}, but with $g_{\mu\nu}$ replaced by $h_{\mu\nu}$), we have
\be\label{det(1-T)DBI}
\det\mathcal G^{(\mu\nu)}=\mathcal G^{-2}h\,,\qquad
\det(\delta_\mu^\nu-\lambda T_\mu{}^\nu)=\mathcal G^{-1}h\,.
\ee
Thus
\be\label{GF=T}
\mathcal G^{\mu\nu}F_{\nu\mu}=\frac 2{\sqrt{\lambda}}\Big( 1-\det(\delta_\mu^\nu-\lambda T_\mu{}^\nu)\Big)
\ee
and the last term in \eqref{ldDBI2} takes the form
\be\label{lastDBI2}
\frac 1{4\lambda\sqrt{\lambda}}\sqrt{\mathcal G g^{-1}}\,\Big(\mathcal G^{\mu\nu}F_{\nu\mu}\Big)=\frac 1{2\lambda^2}\sqrt{\mathcal G g^{-1}}\Big( 1-\det(\delta_\mu^\nu-\lambda T_\mu{}^\nu)\Big).\,
\ee
On the other hand, one can rewrite the second equality in \eqref{GF1} as 
\be\label{GF2}
\sqrt{\lambda}(h^{\kappa\lambda}h^{\alpha\beta}F_{\kappa\alpha}F_{\lambda\beta})\,\mathcal G^{-1}h
=2\mathcal G^{-1}g\,\det (F_{\mu\rho}g^{\rho\nu})\mathcal =\frac 2{\sqrt{\lambda}}\mathcal (1-\mathcal G^{-1}h),
\ee
from which it follows that
\be\label{rootGg}
\sqrt{\mathcal Gg^{-1}}={\sqrt{\lambda}}\frac{\sqrt{\det (F_{\mu\rho}g^{\rho\nu})}}{\sqrt{1-\det(\delta_\mu^\nu-\lambda T_\mu{}^\nu)}}\,.
\ee
Substituting this relation into \eqref{GF1} and the latter into \eqref{ldDBI2} we get the form of the ``flow" equation found in \cite{Blair:2024aqz}
\be\label{Blair2}
\frac 1{\sqrt{-g}}\frac{\partial\mathcal L
_\lambda}{\partial\lambda}=-\frac 1{2\lambda}\tr\, T-\frac 1{2\lambda\sqrt{\lambda}}{\sqrt{-\det (F_{\mu\rho}g^{\rho\nu})}}{\sqrt{\det(\delta_\mu^\nu-\lambda T_\mu{}^\nu)-1}}\,.
\ee
The right-hand side of this equation explicitly depends on the invariant of $F_{\mu\nu}$, so it cannot be regarded as a stress-tensor deformation. As we already mentioned, the reason is that the number of three independent invariants of the fields contained in the $2\times 2$ matrix $\mathcal G_{\mu\nu}$\footnote{One invariant is e.g. $\det(F_{\mu\rho}g^{\rho\nu})$ and other two are $\partial_\mu\Phi^I\partial^\mu\Phi^JG_{IJ}$ and $\det (\partial_\mu\Phi^I\partial_\nu\Phi^JG_{IJ})$.} exceeds the number 2 of independent invariants of the stress-energy tensor $T_{\mu\nu}$.

The number of independent invariants of $\mathcal G_{\mu\nu}$ gets reduced to two when the theory contains only one scalar since $\det \Big(\partial_\mu\Phi\partial_\nu\Phi \,G(\Phi)\Big)=0 $. We will now show that in this case, the DBI action does obey a stress-tensor flow equation.
Now the inverse matrix $h^{\mu\nu}$ has the following form 
\be\label{h-11}
h^{\mu\nu}=g^{\mu\nu}-\lambda\frac{\partial^\mu\Phi\partial^\nu\Phi\, G(\Phi)}{1+\lambda \partial_\rho\Phi\partial^\rho\Phi\, G(\Phi)}=g^{\mu\nu}-\lambda\,\partial^\mu\Phi\partial^\nu\Phi\, G(\Phi)\,gh^{-1}\,,
\ee
and
\be\label{deth1}
h=g(1+\lambda \partial_\rho\Phi\partial^\rho\Phi\, G(\Phi))\,,
\ee
where the indices are raised with the metric $g^{\mu\nu}$. 

Using equations \eqref{G-1}, \eqref{det(1-T)DBI}, \eqref{h-11} and \eqref{deth1} we find that
\be\label{2-trT}
\begin{split}
\sqrt{\mathcal Gg^{-1}}\mathcal G^{\mu\nu}g_{\mu\nu}&=\sqrt{\mathcal Gg^{-1}}\Big(1+\frac 1{1+\lambda \partial_\rho\Phi\partial^\rho\Phi\, G(\Phi)}\Big)\mathcal G^{-1}h \\
&=
\sqrt{\mathcal Gg^{-1}}(1+h^{-1}g) \mathcal G^{-1}h 
=\sqrt{\mathcal Gg^{-1}}(\mathcal G^{-1}h+\mathcal G^{-1}g)=2-\lambda\tr\, T\,
\end{split}
\ee
so that\footnote{For brevity we denoted $\det(\delta_\mu^\nu-\lambda T_\mu{}^\nu)=\det(\mathbf I-\lambda T)$.}
\be\label{sGg-1}
\mathcal Gg^{-1}-\frac{2-\lambda \tr\, T}{\det(\mathbf I-\lambda T)}\sqrt{\mathcal Gg^{-1}}+\frac 1{\det(\mathbf I-\lambda T)}=0\,.
\ee
Solving this quadratic equation for $\sqrt{\mathcal Gg^{-1}}$ we get
\begin{equation}
\label{sGg-1solved}
   \sqrt{\mathcal Gg^{-1}}=\frac{1-\frac \lambda 2\tr\, T\pm\sqrt{\Big(1-\frac \lambda 2\tr\, T\Big)^2-{\det(\mathbf I-\lambda T)}}}{\det(\mathbf I-\lambda T)}= \frac{1-\frac \lambda 2\tr\, T\pm\sqrt{\frac {\tr (T^2)}2-\frac {(\tr\, T)^2} 4}}{\det(\mathbf I-\lambda T)}.
\end{equation}
Notice the appearance here again of the Root-$T\overline T$ operator \eqref{Rg}.
The choice of the sign in front of the square root is correlated with the positive or negative definiteness of $\partial_\mu \Phi\,\partial^\mu \Phi\, G(\Phi)=\pm |\partial_\mu \Phi\,\partial^\mu\Phi\, G(\Phi)|$. 

Substituting the expression \eqref{sGg-1solved} into \eqref{GF=T} and the latter into \eqref{ldDBI2} we get 
\be\label{stflowDBI2}
\begin{aligned}
\frac 1{\sqrt{-g}}\frac{\partial\mathcal L
_\lambda}{\partial\lambda}
={}&-\frac { \tr\, T} {2\lambda}
-
\Bigg(
1-\frac {\lambda \tr\, T} 2
\pm \lambda \sqrt{\frac {\tr (T^2)}2-\frac {(\tr\, T)^2} 4}
\Bigg)
\frac{1-\left(\det(\mathbf I-\lambda T)\right)^{-1}}{ 2\lambda^2}.
\end{aligned}
\ee
Though quite cumbersome, this is a genuine stress-tensor flow equation showing that $2d$ DBI theory with a single scalar field is a $T\overline T$-like deformation of the seed theory action \eqref{DBI_L0} for $d=2$ and $N=1$. Note that $\lambda\to 0$ limit of the right-hand side of this equation is well defined 
\be
\frac 1{\sqrt{-g}}\frac{\partial\mathcal L
_\lambda}{\partial\lambda}
=\frac 12\left (\frac{\tr\, T_0}2\pm \sqrt{\frac {\tr (T_0^2)}2-\frac {(\tr\, T_0)^2} 4}\right)^2+ O(\lambda)\,.
\ee
As a consistency check of \eqref{stflowDBI2} we note that, when $F_{\mu\nu}=0$, we have $\det(\mathbf I-\lambda T)=1$ and the right hand side of \eqref{stflowDBI2} reduces to the NG theory stress-tensor operator $-\tfrac { 1} {2\lambda}\tr\, T=-\tfrac{1}{2}\det T_{\mu}{}^\nu$, while when instead we set $\Phi=0$, the stress-energy tensor satisfies the relation \eqref{eq:det-trace-id-BI} which is equivalent to 
\be
\det(\mathbf I-\lambda T)=\left(1-\frac \lambda 2\tr\, T\right)^2
\ee
and hence \eqref{stflowDBI2} reduces to the $2d$ BI flow equation \eqref{eq:twoflows}.

\subsection{Three dimensions}
As we have already mentioned, this instance of the DBI theory is indeed a stress-tensor deformation for any number of scalar fields \cite{Blair:2024aqz}. As we noted, this property can be explained by the fact that the purely scalar field $3d$ DNG theory and the pure BI theory obey one and the same flow equation \eqref{newbi}, which turns out to be also the flow equation of the $3d$ DBI theory, as we briefly review. As mentioned in \cite{Blair:2024aqz}, this property can also be attributed to the fact that in $d=3$ an Abelian vector gauge field is dual to a massless scalar field.

In $d=3$ the equation \eqref{ldDBI} takes the form 
\begin{equation}\label{DBI_lambda_derivative}
\frac{1}{\sqrt{-g}}\frac{\partial\mathcal{L}_\lambda}{\partial\lambda}
    =
    -\frac{1}{\lambda^2}
    -
    \frac{1}{2 \lambda^2} \sqrt{\mathcal{G}g^{-1}}
    +
    \frac{1}{2\lambda^2} \sqrt{\mathcal{G}g^{-1}} \mathcal{G}^{\mu\nu} g_{\mu\nu}
    +
    \frac{1}{4 \lambda \sqrt{\lambda}} \sqrt{\mathcal{G}g^{-1}} \mathcal{G}^{\mu\nu} F_{\nu\mu}\,.
\end{equation}
One can then show (see Appendix~\ref{app:dbi-derivation}) that 
\begin{equation}\label{DBI_GF}
    \mathcal{G}^{\mu\nu} F_{\nu\mu}
    =
    \frac{2}{\sqrt{\lambda}} 
    \left[1-
    \sqrt{\mathcal{G}^{-1}g} \det \left(\delta^\nu_\mu - \lambda T_\mu{}^\nu\right) 
    \right]\,.
\end{equation}
We now substitute this expression and the one for $\mathcal{G}^{\mu\nu}g_{\mu\nu}$, equation \eqref{GgT}, into the right-hand side of~\eqref{DBI_lambda_derivative}.
All the terms that do not depend on the stress-energy tensor cancel each other, and we find the flow equation 
\begin{equation}\label{final}
\displaystyle   \frac{1}{\sqrt{-g}}\frac{\partial\mathcal{L}_\lambda}{\partial\lambda}
    =
   -\frac{1}{2\lambda}
\left(\tr\, T-\frac{1}{\lambda}\right)
    - 
    \frac{1}{2\lambda^2}
    \det \left(\delta^\nu_\mu - \lambda T_\mu{}^\nu\right)\,,
\end{equation}
which has indeed the same algebraic form as the $d=3$ flow equation for the DNG and BI actions, see \eqref{membraneflowTT},\eqref{membraneflowTT1} and~\eqref{newbi}.

For a generic DBI action in $d\geq 4$ the authors of \cite{Blair:2024aqz} showed that its $\lambda$-derivative  contains terms depending on invariants of $F_{\mu\nu}$. We expect that for a sufficiently small number of scalar fields $N< d$, when the number of independent invariants of $\mathcal G_{\mu\nu}$ do not exceed those of $T_{\mu\nu}$, it may be possible to recast the $\lambda$-derivative of the DBI actions in a form which only depends on $T_{\mu\nu}$ and  $\lambda$. For instance, in $d=4$ the generic matrix $\mathcal G_{\mu\nu}$ has (at least) 10 independent $SO(1,3)$ invariants. If we consider the $4d$ theory with only one scalar field, the number of independent invariants of $\mathcal G_{\mu\nu}$ reduces to 4, which is the same number as the symmetric $T_{\mu\nu}$ has. We leave this issue for a future study.

\subsection{Three-dimensional DBI flow by dimensional reduction}
\label{DBIfromBI}
A possible consistency check of the DBI flow equation \eqref{final} is to obtain it (at least for the case of a single scalar field) by a dimensional reduction of the four-dimensional BI action 
\begin{equation}\label{eq:BI4d}
    \mathcal S_{\lambda}=
    \int \d^4x\,\left(\frac{\sqrt{-g}}{\lambda}-\frac 1\lambda\sqrt{-\det\left(g_{\mu\nu}(x)+\sqrt{\lambda}F_{\mu\nu}\right)}\right)\,,
\end{equation}
where as usual $F_{\mu\nu} = \partial_\mu A_\nu - \partial_\nu A_\mu$. 
We assume that the $x^3$-direction is compactified on a circle with an infinitely small radius. Then only fields independent of the compact direction remain, i.e. 
\begin{equation}\label{assumption_comp}
    \partial_3 A_\mu = 0\qquad \mathrm{for}~ \mu=0,1,2,3\,.
\end{equation}
Accordingly, we split the indices as $\mu=(a,3)$ with $a=0,1,2$.
Consequently, we have for the field strength $F_{a3}=\partial_a A_3 - \partial_3 A_a = \partial_a A_3 = -F_{3a}$.
For the determinant appearing in~\eqref{eq:BI4d} we find 
\begin{equation}
\begin{split}
    \det \left(g_{\mu\nu} + \sqrt{\lambda} F_{\mu\nu}\right)
    &=
    \det
    \begin{pmatrix}
        g_{ab} + \sqrt{\lambda}\, F_{ab} & g_{a3} + \sqrt{\lambda}\, \partial_a A_3 \\
        g_{3b} - \sqrt{\lambda}\, \partial_b A_3 & g_{33}
    \end{pmatrix}=g_{33}   \det (\mathcal{G}_{ab})\,,
\end{split}
\end{equation}
where we introduced
\begin{equation}\label{fiducialMetric_dimreduced}
    \mathcal{G}_{ab}
    =
    g_{ab} 
    +
    \sqrt{\lambda}\, F_{ab}
    +
    \frac{
    \lambda \partial_a A_3 \partial_b A_3
    -g_{a3}g_{b3}
    +
    \sqrt{\lambda}
    \left(
        g_{a3} \partial_b A_3 - \partial_a A_3 g_{b3}
    \right)
    }{g_{33}}\,.
\end{equation}
To dimensionally reduce, we also need to assume that the metric $g_{\mu\nu}$ has no dependence on $x^3$.%
\footnote{More precisely, we assume that $\partial_3$ is an irrotational Killing vector so that the spacetime is foliated in constant-$x^3$ hypersurfaces orthogonal to~$\partial_3$. This is analogous to the definition of a static spacetime.}
In this case, we may assume that the coordinates are such that $\partial_3 g_{\mu\nu}=0$ and $g_{a3}=0$. Then the Lagrangian in~\eqref{eq:BI4d} is
\begin{equation}
    \mathcal{L}_{\lambda}
    =
    \frac{\sqrt{-g}}{\lambda}
    -
    \frac{1}{\lambda}
    \sqrt{-g_{33}\cdot \det \left(
    g_{ab} + \sqrt{\lambda}\, F_{ab}
    +
    \lambda\, \partial_a A_3 \partial_b A_3\,g^{33} 
    \right)} \,.
\end{equation}

We would now like to see what form the flow equation \eqref{p=3} takes upon the reduction from $d=4$ to $d=3$.
In the four-dimensional case, the deformation operator is simply proportional to the trace $T^{\mu\nu} g_{\mu\nu}$.
With the abbreviation $g = \det (g_{\mu\nu}) = g_{33} \det (g_{ab}) \equiv g_{33} \tilde{g}$ one has
\begin{subequations}
    \begin{align}
            T^{ab}
        &=
        \frac{1}{\lambda}
        \left[
            g^{ab}
            -
            \sqrt{\mathcal{G}\tilde{g}^{-1} }\,
            \mathcal{G}^{(ab)}
        \right] \, ,\\
        T^{33}
        &=
        \frac{1}{\lambda}
        \left[
            g^{33}
            -
            g^{33}
            \sqrt{\mathcal{G}\tilde{g}^{-1}}
            +
           \lambda (g^{33})^2 
            \sqrt{\mathcal{G}\tilde{g}^{-1}}\,            
            \mathcal{G}^{ab} \,\partial_a A_3 \partial_b A_3
        \right] \, .
    \end{align}
\end{subequations}
If the operations of fixing the fields as $\partial_3 A_\mu = 0$ and performing the derivatives $\partial_\lambda$ and $\partial_{g_{\mu\nu}}$ (for the stress-energy tensor) commute, as they should, the three-dimensional theory inherits the flow equation \eqref{p=3}.
Of course, one should keep in mind that the trace of the three-dimensional stress-energy tensor, $T^{ab}$, does not coincide with the operator appearing on the right-hand side of \eqref{p=3}, which is constructed from the original four-dimensional tensor $T^{\mu\nu}$. Indeed, one finds that the inherited flow equation reads
\begin{equation}
\begin{split}
        \frac{1}{\sqrt{-\tilde{g}\,g_{33}}}
    \frac{\partial\mathcal{L}_\lambda}{\partial\lambda}
        &=
        -\frac{1}{4\lambda} T^{\mu\nu} g_{\mu\nu} 
        =
        -\frac{1}{4\lambda} \left( 
        T^{ab} g_{ab}
        +
        T^{33} g_{33}
        \right)\\
        &=
        -\frac{1}{4\lambda} T^{ab}g_{ab}
        -
        \frac{1}{4 \lambda^2}
        +
        \frac{1}{4\lambda^2}\sqrt{\mathcal{G}\tilde{g}^{-1}}
        -
        \frac{1}{4\lambda}
        \sqrt{\mathcal{G}\tilde{g}^{-1}}\,
        \mathcal{G}^{ab}
        \partial_a A_3 \partial_b A_3 g^{33}\,.
\end{split}
\end{equation}
The last line is the familiar expression for the $\lambda$-derivative of the DBI Lagrangian, as given in \eqref{DBI_lambda_derivative}.
To see this, the field strength should be substituted by $F_{ab} =( \mathcal{G}_{ab} - g_{ab} - \lambda \,\partial_a \Phi \partial_b\Phi \, G(\Phi) )/\sqrt{\lambda}$, while $A_3=\Phi$ and $g^{33}=G(\Phi)$.
In the previous section, we showed that \eqref{DBI_lambda_derivative} can be expressed purely in terms of the three-dimensional stress-energy tensor, yielding the flow equation \eqref{final}.
This demonstrates that the three-dimensional flow equation of single-scalar DBI \eqref{final} can be obtained by the dimensional reduction of the four-dimensional flow equation of BI theory \eqref{p=3}.

\section{Conclusion and outlook}\label{conclusions}

In this work, we have considered some possible generalisations of $T\overline{T}$ flows to higher dimensions. In $d=2$, $T\overline{T}$ flows have a remarkable number of features, and it is quite likely that only some may be preserved by higher-dimensional generalisations. The original results of this paper include the following ones.

Using the approach of~\cite{Seibold:2023zkz}, we attempted to construct $d=3$ $T\overline{T}$-like deformations by dimensional uplift, using as a seed theory the three-dimensional massless scalar one. The result is non-local and can be written as an expansion in the deformation parameter
\begin{equation}
\label{eq:nonisotropicaction}
\mathcal{S}_\lambda
= \int \mathrm{d}^2x \!\!\int\limits_{0}^{2\pi R}\!\! \mathrm{d}y
\Bigg(\mathcal{L}_0(x,y)+\frac{\lambda}{2}\!\!\int\limits_{0}^{2\pi R}\!\!\d y'\Bigg(\mathcal{O}_1(x,y;y')+\frac{\lambda}{2}\!\!\int\limits_{0}^{2\pi R}\!\!\d y''\Big(\mathcal{O}_2(x,y;y',y'')+\dots\Big)\Bigg)\Bigg),
\end{equation}
where the seed Lagrangian is simply $\mathcal{L}_0=-\frac{1}{2}\partial_\mu\Phi\partial^\mu\Phi$. The first few corrections can be written down quite explicitly, see \eqref{LO}, \eqref{NLO}, and \eqref{NNLO}, but it is hard to see the pattern of subsequent terms. This, together with the non-local and non-isotropic form of the action, makes it difficult to understand the underlying physics.

A different approach, more naturally tailored to higher dimensions (but perhaps less directly related to standard $T\overline{T}$ deformations) is to start from notable actions, depending on a parameter, and try to interpret them as  $T\overline{T}$-like flows (i.e., those driven by the stress-energy tensor).
For the Dirac--Nambu--Goto-type theories (where the parameter $\lambda$ is the inverse tension) we reviewed the derivation of the stress-tensor flow equation \eqref{eq:flowEqnoV},
\be
\frac 1{\sqrt{-g}}\frac{\partial \mathcal L_\lambda}{\partial \lambda}
=-\frac 1{2\lambda}\left(\tr\, T+\frac{2-d}\lambda\right)+\frac{2-d}{2\lambda^2}\left(\det\left(\delta_\mu^\nu-\lambda T_{\mu}{}^{\nu}\right)\right)^{\frac 1{d-2}},
\ee originally found in \cite{Blair:2024aqz}. We also obtained its solution when the seed action is that of a sigma model with a potential $V(\Phi)$, see equation~\eqref{f=k=1}. This result has not appeared in the literature before. For the $d$-dimensional action with one scalar field and potential $V(\Phi)$, whose flow is governed by equation \eqref{Rgflow} we recovered the result~\eqref{STTVg} of~\cite{Babaei-Aghbolagh:2024hti}.

It is worth noting that, unlike~\eqref{eq:nonisotropicaction}, the deformed actions~\eqref{f=k=1} and~\eqref{STTVg} enjoy $d$-dimensional Poincar\`e invariance. Hence, arguments \textit{\`a la} Coleman--Mandula~\cite{Coleman:1967ad} suggest that the resulting relativistic dynamics would not be integrable if $d>2$, at least in terms of the scattering matrix.

For $d$-dimensional Born--Infeld-like theories we  obtained, for the first time, the generic stress-tensor flow equation \eqref{BIflowT}, namely
\begin{equation}
\displaystyle   \frac{1}{\sqrt{-g}}
    \frac {\partial\mathcal{L}_\lambda}{\partial\lambda} 
    =
    \frac{d-4}{4\, \lambda^2}
    -
    \frac{\tr\,T}{4 \,\lambda}
    +
    \frac{4-d}{4\, \lambda^2}
    \left(
        \det \left(\delta_\mu^\nu - \lambda T_\mu^{~\nu}\right)
     \right)^{\frac{1}{d-4}}.
\end{equation}

In the case of Dirac--Born--Infeld-type theories, we focused on $d=2$ and $d=3$. In $d=2$ we derived, for the first time, the stress-tensor flow equation for the DBI action with a single scalar field, equation
\eqref{stflowDBI2},
\be
\begin{aligned}
\frac 1{\sqrt{-g}}\frac{\partial\mathcal L
_\lambda}{\partial\lambda}
={}&-\frac { \tr\, T} {2\lambda}
-
\Bigg(
1-\frac {\lambda \tr\, T} 2
\pm \lambda \sqrt{\frac {\tr (T^2)}2-\frac {(\tr\, T)^2} 4}
\Bigg)
\frac{1-\left(\det(\mathbf I-\lambda T)\right)^{-1}}{ 2\lambda^2},
\end{aligned}
\ee
which contains the Root-$T\overline{T}$ operator in the ($d$-dependent) form introduced in~\cite{Babaei-Aghbolagh:2022uij,Ferko:2022iru,Conti:2022egv}. In $d=3$, we showed how the DBI stress-energy flow derived in \cite{Blair:2024aqz} can be obtained from the dimensional reduction of the $d=4$ BI flow.

Our results further elucidate, from the stress-tensor deformation perspectives, the structure of classes of non-liner theories in higher dimension which are of interest, in particular in the context of String Theory. They indicate possible obstructions on realizing generic p-brane DBI actions as $T\overline T$-like deformations thus questioning the applicability of these techniques to study field-theoretical and physical properties of the latter.

In this respect, there are several natural questions related to this work which can be pursued elsewhere, starting from extending our results for DBI flows to dimensions $d\geq 4$, as well as to DBI theories that include couplings to two- and higher-form gauge fields in the bulk, like in the case of the Dirichlet $p$-branes of string theory.
In particular, one would like to see (at least for certain values of $p$ and $d$) whether the deforming operator is expressible exclusively in terms of scalar invariants of the stress-energy tensor. As already noted, the $d=4$ case with a single scalar field appears here especially promising.
It would also be interesting to further explore possible links between non-Abelian
generalizations of the BI-type theory to
stress-tensor deformations of Yang-Mills theories, see e.g.~\cite{Ireland:2019vvj,Brennan:2019azg,Griguolo:2022hek,Ferko:2024yua}.

It would be interesting to explore possible connections with higher-dimensional (massive) gravity theories~\cite{Conti:2022egv,Floss:2023nod,Morone:2024ffm, Tsolakidis:2024wut, Nix:2025plr}, where stress-tensor deformations may play an analogous structural role, as well as to study the relation of these flows to higher-dimensional theories with non-linearly realised supersymmetry \cite{Cribiori:2019xzp, Ferko:2019oyv, Ferko:2023ruw, Ferko:2023sps}.

Unlike what happens in two dimensions~\cite{Frolov:2019nrr, Sfondrini:2019smd}, it is not {\it a priori} clear whether these $T\overline T$-like deformations may be interpreted as emerging from the action of the membrane ~\cite{Howe:1977hp, Howe:1977ut} in a ``uniform'' light-cone gauge, and it would be interesting to understand if this is the case. 
In this respect, it is worth noting that in contrast to what happens to the string, the membrane remains interacting in the standard light-cone gauge~\cite{deWit:1988wri} even in a flat Minkowski background.

Another approach which has been proven useful to better understand $T\overline{T}$ deformations in two dimensions, is to interpret them as deformations of a background metric, see e.g.~\cite{Conti:2022egv,Babaei-Aghbolagh:2024hti,Li:2025lpa}. These works have already provided such an interpretation for the case of the single scalar field considered in section \ref{single}. It would be interesting to see explicitly how other higher-dimensional flows considered here emerge within these geometrical settings.

\paragraph{Acknowledgements.} 
We are grateful to Hossein Babaei-Aghbolagh, Andrea Cavagli\`a, Christian Ferko, Sergei Kuzenko, Gregory Korchemsky, Alexia Nix, Niels A.~Obers, Fiona K.~Seibold, Gabriele Tartaglino-Mazzucchelli, Roberto Tateo, Evangelos Tsolakidis, and Ziqi Yan for useful discussions and comments on the manuscript.
This work was supported by the CARIPARO Foundation under grant No.~68079. Work of DS was supported in part by the Australian Research Council, project DP230101629, the MCI, AEI, FEDER (UE) grant PID2021-125700NB-C21 ``Gravity, Supergravity and Superstrings'' (GRASS), and the Basque Government Grant IT1628-22. D.S. also thanks Physics's Department, the University of Western Australia (Perth) and School of Mathematics and Physics, the University of Queensland (Brisbane) for kind hospitality during the final stages of this work.

\appendix

\section{Derivation of the DBI flow in three dimensions}
\label{app:dbi-derivation}
The goal of the appendix is to provide a derivation of \eqref{DBI_GF} by calculating $\mathcal{G}^{\mu\nu} F_{\nu\mu}$.
Therefore, we have to determine the inverse tensor $\mathcal{G}^{\mu\nu}$ of $\mathcal G_{\mu\nu}$.
To do so, we make three instrumental observations.

First, we notice that $\mathcal{G}_{\mu\nu}$ is composed of 
an induced metric
\be
h_{\mu\nu} \equiv g_{\mu\nu} + \lambda\, \partial_\mu \Phi^I \partial_\nu \Phi^J\, G_{IJ}(\Phi)\,.
\ee together with the antisymmetric field strength
\be\label{mG}
\mathcal{G}_{\mu\nu} = h_{\mu\nu} + \sqrt{\lambda}\, F_{\mu\nu}
\ee
We denote the determinant of the induced metric as $h \equiv \det (h_{\mu\nu})$.
To recover the case of the three-dimensional brane, section \ref{NGexamples}, we set $F_{\mu\nu} = 0$ and have $h_{\mu\nu} = \mathcal{G}_{\mu\nu}$. 
To specialize to the three-dimensional BI-type model, section \ref{BIexamplesMoreThan2d}, we switch off the scalar fields and have $h_{\mu\nu} = g_{\mu\nu}$.
We encountered the tensor $h_{\mu\nu}$ in section \ref{DNG} as the induced metric of  scalar fields $\Phi^I(x)$ (denoted by $\mathcal{G}_{\mu\nu}$ therein). 

Second, to compute the determinant of the stress-energy tensor, we make use of the following relations
\begin{subequations}
\begin{align}
    \mathcal{G}^{(\mu\nu)}
    &=
    \mathcal{G}^{\mu\rho}
    \mathcal{G}_{(\rho\sigma)}
    (\mathcal{G}^{-\mathrm{T}})^{\sigma\nu}
    =
    \mathcal{G}^{\mu\rho}
    h_{\rho\sigma}
    (\mathcal{G}^{-\mathrm{T}})^{\sigma\nu}\, ,\label{DBI_Inverse_G_ito_G} \\
    \det\,\mathcal{G}^{(\mu\nu)}
    &=
   ( \det\, \mathcal{G}^{\mu\nu})^2 \cdot
    h\,,
    \label{invGsym_det} \\
    \det \,(\delta^\nu_\mu - \lambda T_{\mu}{}^{\nu})
    &=
    h\, g^{-1} \sqrt{\mathcal{G}^{-1}g}
    ~.\label{DBI_det_1-lT}
\end{align}
\end{subequations}
The first equation \eqref{DBI_Inverse_G_ito_G} can be proven by analogy with the BI case \eqref{BI_Gmatinvsym_relation}.

Third, to facilitate the computation, we introduced the dual field strength vector 
\be f^\mu \equiv \frac{1}{2\sqrt{-h}}\varepsilon^{\mu\nu\rho}F_{\nu\rho}\ee 
and its square 
\be f^2 \equiv h_{\mu\nu} f^\mu f^\nu = f^\mu f_\mu\,.\ee
Inversely we have 
\be F_{\mu\nu} = -\sqrt{-h}\,\varepsilon_{\mu\nu\rho}f^\rho\,,\ee
which immediately implies 
\be f^\mu F_{\mu\nu}\equiv 0\,.\ee
The relations for the three-dimensional Levi--Civita symbol in Minkowskian signature read
\begin{equation}
    \begin{array}{ccc}
        \varepsilon^{\mu \nu \rho} \varepsilon_{\mu \sigma \tau}
        =
        - 2 \delta^{[\nu}_\sigma \delta_\tau^{\rho]}
        =
        \delta^{\rho}_\sigma \delta_\tau^{\nu} - \delta^{\nu}_\sigma \delta_\tau^{\rho} 
        & 
        \quad\mathrm{and} \quad 
        &
        \varepsilon^{\mu \nu \rho} \varepsilon_{\mu \nu \sigma}
        =
        -2 \delta_\sigma^\rho ~.\\
    \end{array}
\end{equation}
This implies 
\begin{equation}
\begin{array}{ccc}
    F^{\mu\nu}F_{\nu \rho} = f^2 \delta^\mu_\rho - f_\rho f^\mu 
    &
    \mathrm{and}
    &
    F^{\mu \nu}F_{\nu \mu} = 2 f^2 \, ,
\end{array} 
\end{equation} 
where mind that the indices here are raised with the metric $h^{\mu\nu}$ and lowered with $h_{\mu\nu}$.

The above identities allow us to easily guess the form of the inverse matrix, which is
\be\label{imG}
\mathcal G^{\mu\nu}=\frac 1{1-\lambda f^2}(h^{\mu\nu}-\sqrt{\lambda}F^{\mu\nu}-\lambda f^\mu f^\nu)
\ee
Hence we get 
\be\label{GF}
\mathcal G^{\mu\nu}F_{\nu\mu}
=-\frac{2\sqrt{\lambda} f^2}{1-\lambda f^2}
=
\frac{2}{\sqrt{\lambda}}
(1-\mathcal{G}^{-1}h)
=
\frac{2}{\sqrt{\lambda}} 
\left[
    1
    -
    \sqrt{\mathcal{G}^{-1}g} \,\det \left(\delta^\nu_\mu - \lambda T_\mu{}^\nu\right)
\right]\,,
\ee
where to get the second equality, we used the expression for the determinant of $G_{\mu\nu}$
\begin{equation}
    \mathcal{G}
    =
    h \,(1-\lambda f^2)\,.
\end{equation}
The last equality in \eqref{GF} is provided by \eqref{DBI_det_1-lT}. It yields the equation \eqref{DBI_GF} in the main text.

\bibliographystyle{JHEP}
\bibliography{references}

\end{document}